\documentclass{amsart}
\usepackage[scaled]{helvet}
\usepackage[utf8]{inputenc}

\usepackage{amsmath}
\usepackage{amsthm}
\usepackage{setspace}
\usepackage{appendix}
\usepackage{ifthen}
\usepackage{url}
\usepackage{amssymb}
\usepackage{comment}
\usepackage{float}
\PassOptionsToPackage{hyphens}{url}
\numberwithin{equation}{section}
\usepackage{pdfpages}
\newtheoremstyle{break}
  {\topsep}{\topsep}%
  {\upshape}{}%
  {\bfseries}{}%
  {\newline}{}%
\theoremstyle{break}

\newtheorem*{lemma*}{Lemma}

\newtheorem*{coro*}{Corollary}

\usepackage[a4paper, total={6in, 8in}]{geometry}
\begin{document}
\title{Thinned COE random matrix models for DNA replication}
\author{Huw Day and Nina C. Snaith}
\begin{abstract}

   This paper details an observation that for more primitive organisms, such as some yeasts, the statistical distribution of the origins of replication sometimes looks remarkably like the distribution of eigenvalues from the Circular Orthogonal Ensemble (COE) of random matrices.  This does not hold for more complex organisms, but a uniform thinning of the COE eigenvalues (which interpolates between the COE and uncorrelated, Poisson statistics)  gives a platform to investigate characteristics of replication origin distribution in other species where data is available. 
\end{abstract}

\maketitle

\section{Introduction to Eukaryotic DNA Replication}
\label{sect:intro}

Before a cell divides, the DNA must replicate.  The DNA molecule is in essence a linear sequence of pairs of chemical bases, each pair forming the rung of a ladder-like structure.  Replication commences at hundreds of origins along this sequence of base pairs, and progresses in two directions from each origin. Modeling the DNA molecule with a line, the replication origins are points on this line, and it is their spatial distribution that we are interested in.  It seems reasonable that origins should not cluster too closely, as the expanding replication forks would almost immediately meet and coalesce, which is an inefficient use of resources.  On the other hand if there are large gaps between origins then there is a risk that the replication process could go wrong as it spans that gap.  

\begin{figure}
    \centering
    \includegraphics{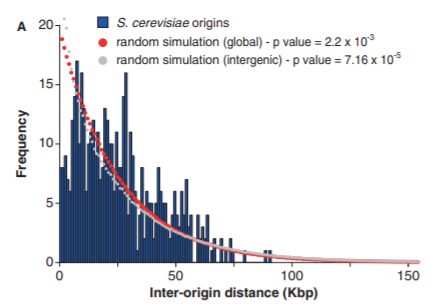}
    \caption{This  is Figure 3A from \cite{Newman}. Original caption: ``Inter-origin spacings in the S. cerevisiae genome. (A) Interorigin spacings in S. cerevisiae were calculated and assigned to different 1 kb bins. The frequency of origins in each bin is shown. Red dots: mean origin separation in a computer simulation where the same number of origins were placed at random on the whole S. cerevisiae genome. Grey dots: mean origin separation in a computer simulation where the same number of origins were placed at random only in the intergenic regions of the S. cerevisiae genome"}
    \label{fig:NewmanFig}
\end{figure}

We see this behaviour in the histogram of spacings between neighbouring origins of a yeast, S. cerevisiae, shown in Figure \ref{fig:NewmanFig} (which is Figure 3A from \cite{Newman}). In this figure, we note that the spacings between the replication origins  appear to exhibit some sort of local repulsion (two origins are unlikely to be close together, which is made evident by the blue histogram being lower close to an inter-origin distance of $0$) and also that two origins are  unlikely to be very far apart, which is made evident by the decrease of the histogram as the inter-origin distance gets larger. This implies that the positions of the points are correlated, as the picture  can be seen to be very different from Poisson /exponential spacings of completely random, uncorrelated points (represented by the red line in Figure \ref{fig:NewmanFig}).

We should note that DNA replication is a hugely complicated process, with many factors influencing the position of replication origins (see \cite{kn:hu_stillman} for a review of recent literature).  There are also variations in the replication process from organism to organism and here we are reducing it to  just looking at the positions at which replication starts, without taking account of the biological process.  In a companion paper \cite{kn:DaySnaith2} we investigate a stochastic model which goes a little further into the process, modeling the expanding replication forks and the consequences of the fact that they may not all start replication at the same time. 

\section{Random Matrix Theory}
\label{sect:RMT}

\begin{figure}[H]
    \centering
    \includegraphics[scale=0.5]{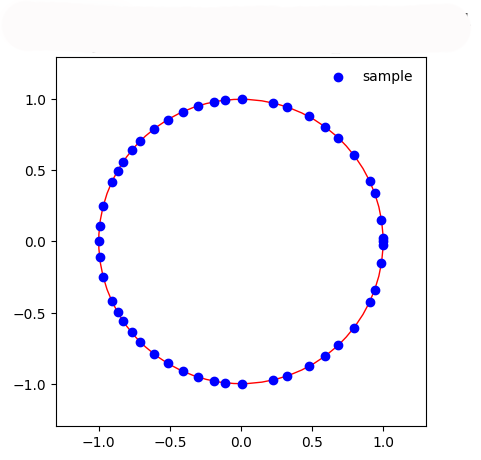}\includegraphics[scale=0.5]{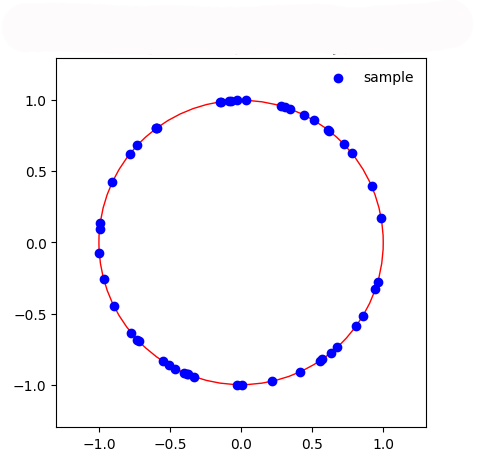}
    \caption{On the left is a visualisation of the eigenvalues of a typical $50\times 50$ COE matrix. Notice here that points are far less likely to cluster or have large spacings than in the figure on the right, which features 50 random, uncorrelated points.}
    \label{fig:coe}
\end{figure}

Studying the eigenvalue distribution of Hermitian or unitary random matrices also amounts to describing the distribution of points on a line. In the case of eigenvalues of standard ensembles of random matrices, the eigenvalue distribution is very distinctive, as the points display repulsion (that is, they tend not to occur very close together) and they tend not to leave large spaces, unlike uncorrelated points.  In the most basic definition of a random matrix, the elements are filled with some type of random variables, possibly respecting some overall symmetry of the matrix.  It is the Jacobian of the change of variables from the matrix elements to the eigenvalue variables that results in repulsion between eigenvalues.  In this paper we will be interested in the Circular Orthogonal Ensemble (COE) of random matrix theory, which consists of all symmetric unitary matrices of a given dimension, endowed with a natural measure that allows us to speak of a ``random" COE matrix (see \cite{forresterbook}, Definition 2.2.3,  or \cite{mehta}, Theorem 10.1.13, for precise definitions).  Eigenvalues of COE matrices display repulsion (see Figure \ref{fig:coe}).

Random matrix statistics are more traditionally associated with modelling systems in various branches of physics which are outlined in detail in \cite{forresterbook}, \cite{mehta} and \cite{taormt}. Random matrices also have applications in material science \cite{Weaver}, signal detecting \cite{SigDet1}, \cite{SigDet2}, \cite{SigDet3} and \cite{SigDet4}, wireless communication \cite{WireCom} and finance \cite{Finance}. Many of these systems have  matrices and eigenvalues associated with them, making it relatively intuitive to attempt to model them with random matrices. In the case of origins of replication, there is no apparent matrix present, but there is precedent for random matrix statistics being observed in systems where there is no such inherent matrix present. Perhaps one of the better known models is the Buses of Cuernavaca system \cite{warchol2018buses,kn:krbseb00,Bus}, where the times between passing buses were shown to display the same statistical behaviour as spaces between eigenvalues of random matrices. In that case there is not only statistical evidence, but also analytical calculations on a stochastic model that simulates the process in question.

There are many ensembles of random matrices, but early experimentation by the students thanked in the Acknowledgments suggested that the COE was the best fit to replication origin data, at least for the more primitive organisms, yeasts, for which data was available first.  For $2\times 2$ COE matrices, the distribution of spacings between neighbouring eigenvalues, a statistic called the nearest neighbour spacing distribution,  is given by Wigner's surmise (see \cite{mehta}),

\begin{equation}
\label{Wigner}
p(s)=\frac{\pi s}{2} \exp \left(-\frac{\pi s^2}{4}\right),
\end{equation}
and it is well-established  that this is also an excellent approximation for COE matrices of larger size, so we will start by comparing this curve to the distribution of spacings between replication origins.

\begin{figure}
    \centering
    \includegraphics[scale=0.5]{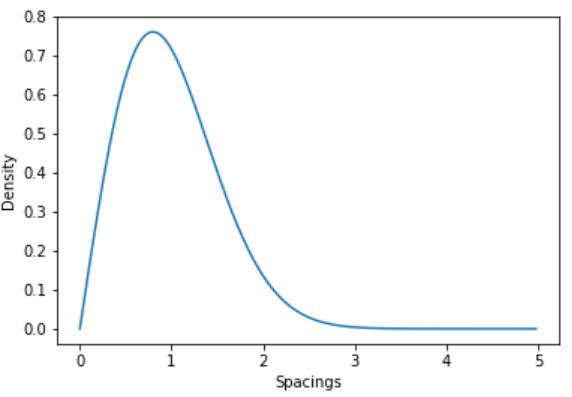}
    \includegraphics[scale=0.5]{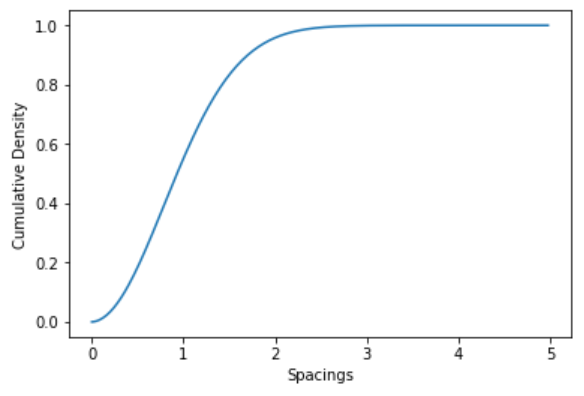}
    \caption{The probability density function (left) and cumulative density function (right) for the distribution of spacings between $2\times2$ COE eigenvalues normalised to have unit average spacing: Wigner's surmise \(p(s)\) from \eqref{Wigner}.  The height of this second graph at a particular spacing represents the proportion of spacings which are less than that particular spacing.}
    \label{fig:Wig}
\end{figure}

The probability density in \eqref{Wigner} is scaled so that   the mean spacing is $1$. Re-scaling is a common technique in random matrix theory in order to compare statistics between different datasets. We will therefore be re-scaling the replication origin spacing data so that the mean of the dataset is $1$. Thus Wigner's surmise is an approximation to the probability density function for the spacing between re-scaled eigenvalues of $N\times N$ COE matrices, and it is plotted in Figure \ref{fig:Wig}.

\section{Genetics Dataset Overview} \label{sect:data}

We will consider a variety of organisms whose replication origin spacings show different statistical distributions. Replication origins have width, as they are many base pairs long, but for our purposes we will model them as points of zero width, with location at the average between the left most part of the origin and the right most part of the origin. This is the standard approach  in the literature (see \cite{Newman} and \cite{Overcook}), but does come with drawbacks which we will look at later.

Each dataset looks like a table with three columns of importance: Chromosome Number, Replication Origin Start, Replication Origin End. The position of replication origins is measured in base pairs (bp) from one end of the chromosome. One base pair is about $0.34$ nanometers. From there, we calculate the Replication Origin Midpoint (again in base pairs) by just taking the average of the Origin start and End and add this as a fourth column to our table. Below is a snapshot (just four entries, two from the end of the first chromosome and two from the start of the second chromosome) of data from the yeast strain K lactis (data taken from \cite{Klactis}):

\begin{table}[H]
\centering
\begin{tabular}{||c | c | c | c ||} 
 \hline
 Chromosome Number & RO Start (bp) & RO End (bp) & \textit{RO Midpoint} (bp) \\ [0.5ex] 
 \hline\hline
 
 1 & 944686 & 944803 & 944744.5\\ 
 \hline
 1 & 1028681 & 1029091 & 1028886 \\ 
 \hline
 2 & 59470 & 60081 & 59775.5 \\ 
 \hline
 2 & 141706 & 142185 & 141945.5 \\
 \hline

\end{tabular}
\caption{An example table of what a snapshot of the raw data looks like to illustrate how we locate replication origin spacings.}
\label{table:snapshot}
\end{table}

To obtain data on inter-origin spacings, we take the difference of neighbouring midpoints provided they are on the same chromosome. From Table \ref{table:snapshot} we would end up with two spacings, one from the pair of origins on chromosome 1 and the other from the pair of origins on chromosome 2: $1028886-944744.5=84141.5$ and $141945.5-59775.5=82170$.

This process is done with all origins and all chromosomes from the data set of a particular organism  to produce a list of spacings between neighbouring replication origins, with the data pooled from all chromosomes so as to maximise the quantity of data. Then the list of spacings is re-scaled so they have mean spacing $1$ to allow for better comparison between other datasets and the Wigner surmise.  

A histogram is created from all the re-scaled spacings of a given organism and it is normalised so that the area of the histogram is 1. This has the effect of turning a histogram into something that approximates a probability density function. This gives us a common reference frame to compare all of our frequency plots.

We also produce cumulative plots from these histograms.  These can then be compared to  the cumulative form of Wigner's surmise, as seen in Figure \ref{fig:Wig}.

Our measure of how well curves match is the Root Mean Square Error (RMSE) between the plot of the data and the respective model (e.g. Wigner's surmise or an exponential random variable). We usually calculate the RMSE on the cumulative plot where some of the random scatter is averaged out. To calculate the RMSE in a comparision of, say, DNA data versus Wigner's surmise, fix $n$ points on the horizontal axis. For both the DNA data and Wigner's surmise, we have the  heights of their cumulative curves at these $n$ points:  \(\{d_{1}, \ldots, d_{n}\}\) and \(\{w_{1}, \ldots, w_{n}\}\) respectively.   We can calculate the RMSE between the two plots by considering the squared difference between each data value and the model value, summing them, scaling this by the number of data points and then square rooting:

\begin{equation}
RMSE:=\sqrt{\frac{1}{n} \sum_{i=1}^{n}(d_{i}-w_{i})^2}.
\end{equation}

The RMSE will always be non-negative, and the closer the value is to $0$, the smaller the error and the better the fit between our two distributions.

\section{Comparison with COE statistics}
\label{sect:COE}

\begin{figure}[H]
    \centering
    \includegraphics[scale=0.3]{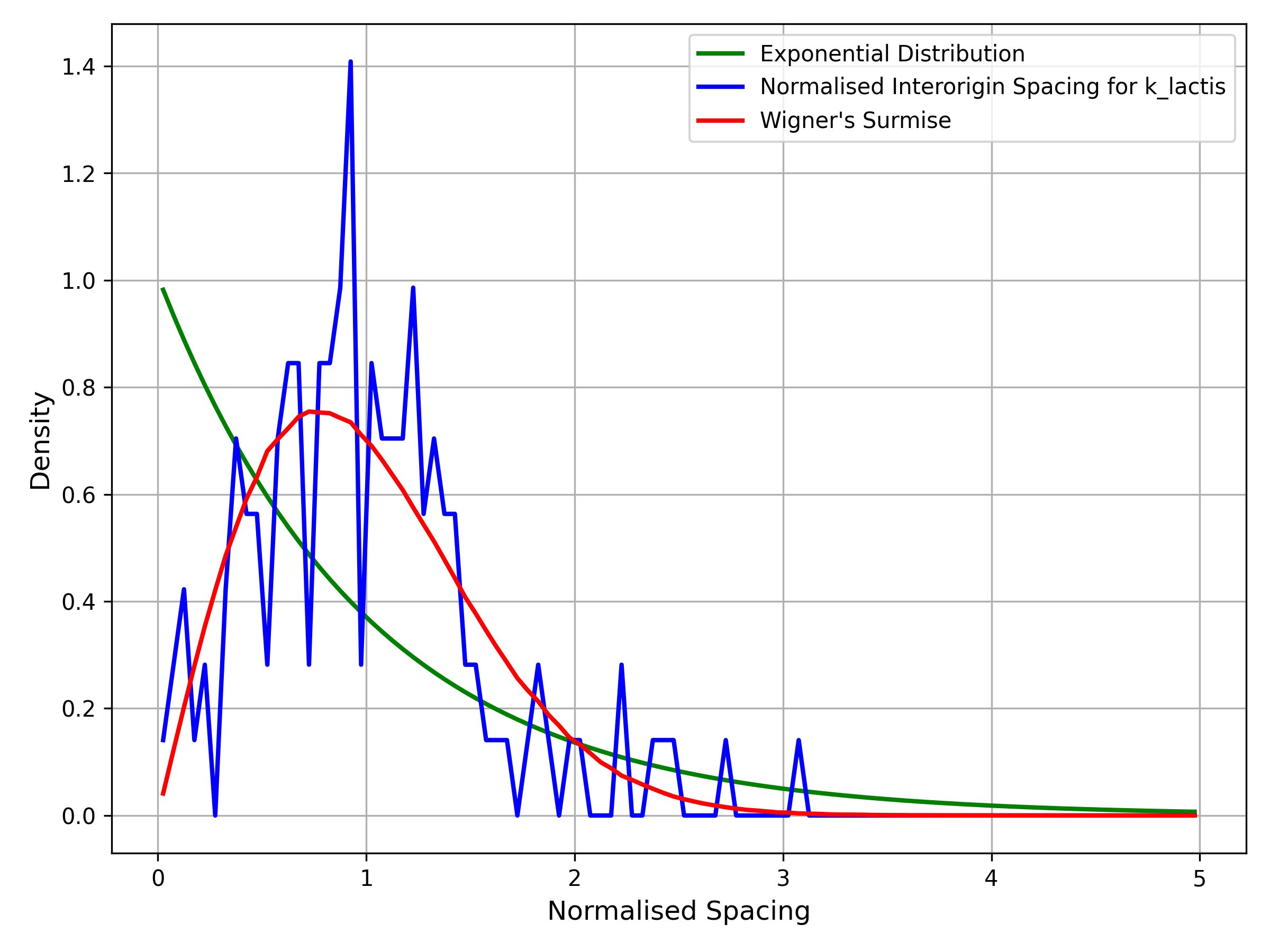}\includegraphics[scale=0.3]{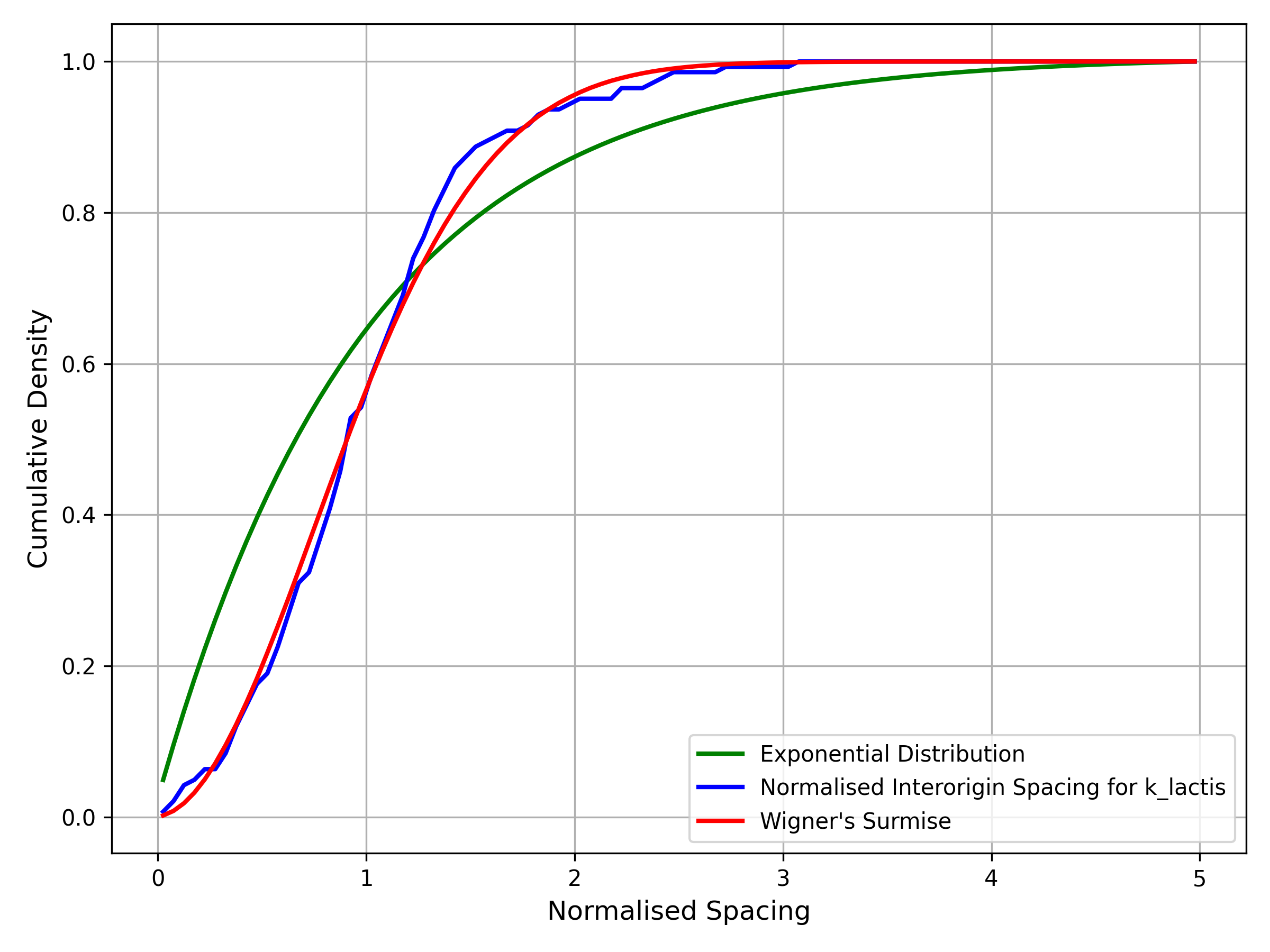}
    \caption{Left: Histogram of re-scaled spacings between midpoints of adjacent replication origins from the yeast strain Kluyveromyces lactis (or K lactis), data taken from \cite{Klactis}, with Wigner's surmise and exponential distribution for comparison. Right: Cumulative distribution of the same data.  Total Number of Spacings: 142. Number of Chromosomes: 6.}
    \label{fig:KlactisWig}
\end{figure}

\begin{figure}[H]
    \centering
    \includegraphics[scale=0.3]{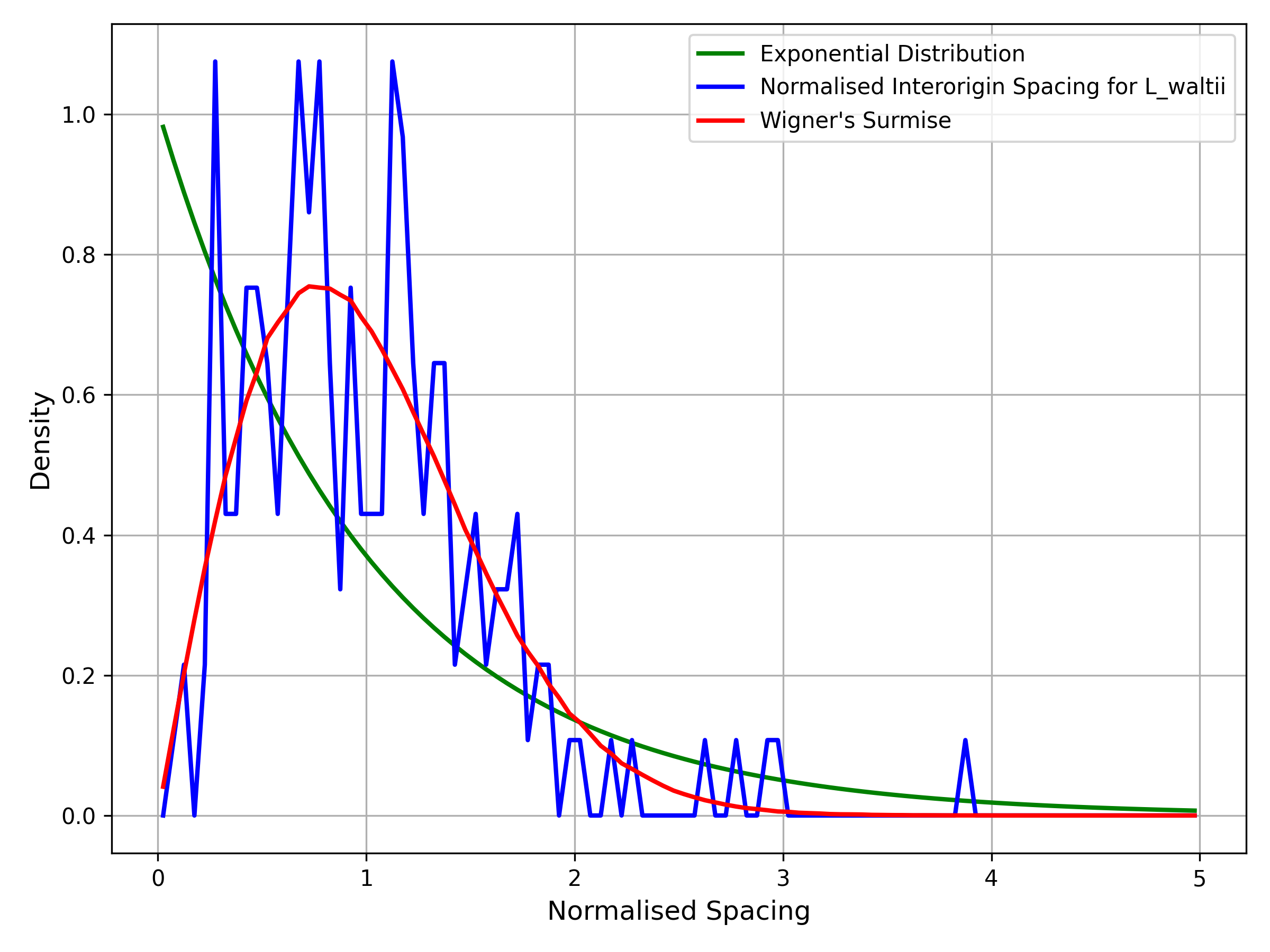} \includegraphics[scale=0.3]{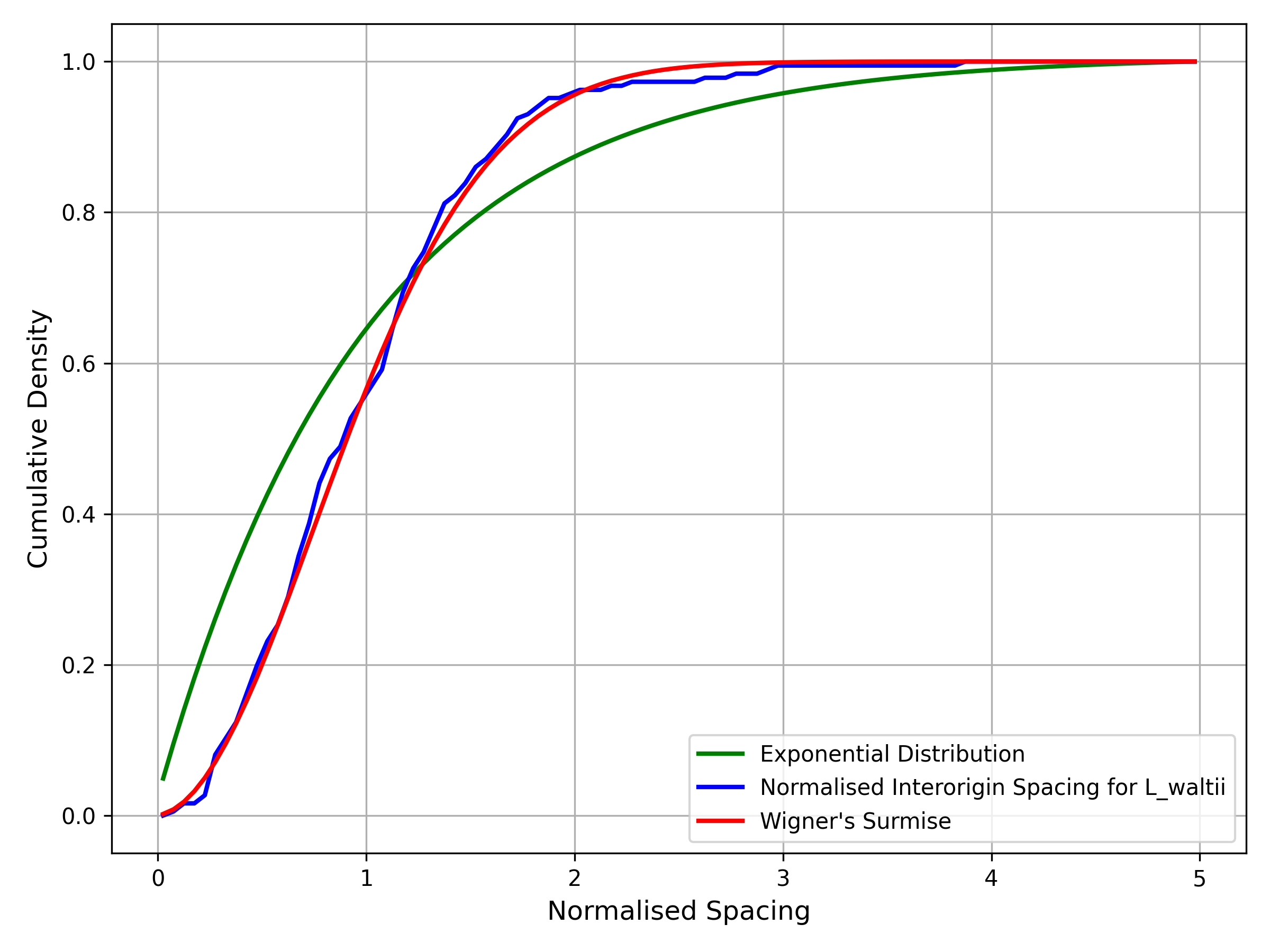}
    \caption{Left: Histogram of re-scaled spacings between midpoints of adjacent replication origins from the yeast strain Lachancea waltii (or L waltii), data taken from \cite{lwaltti2}, with Wigner's surmise and exponential distribution for comparison. Right: Cumulative distribution of the same data.  Total Number of Spacings: 186. Number of Chromosomes: 8.}
    \label{fig:lwatiiWig}
\end{figure}

\begin{figure}[H]
    \centering
    \includegraphics[scale=0.3]{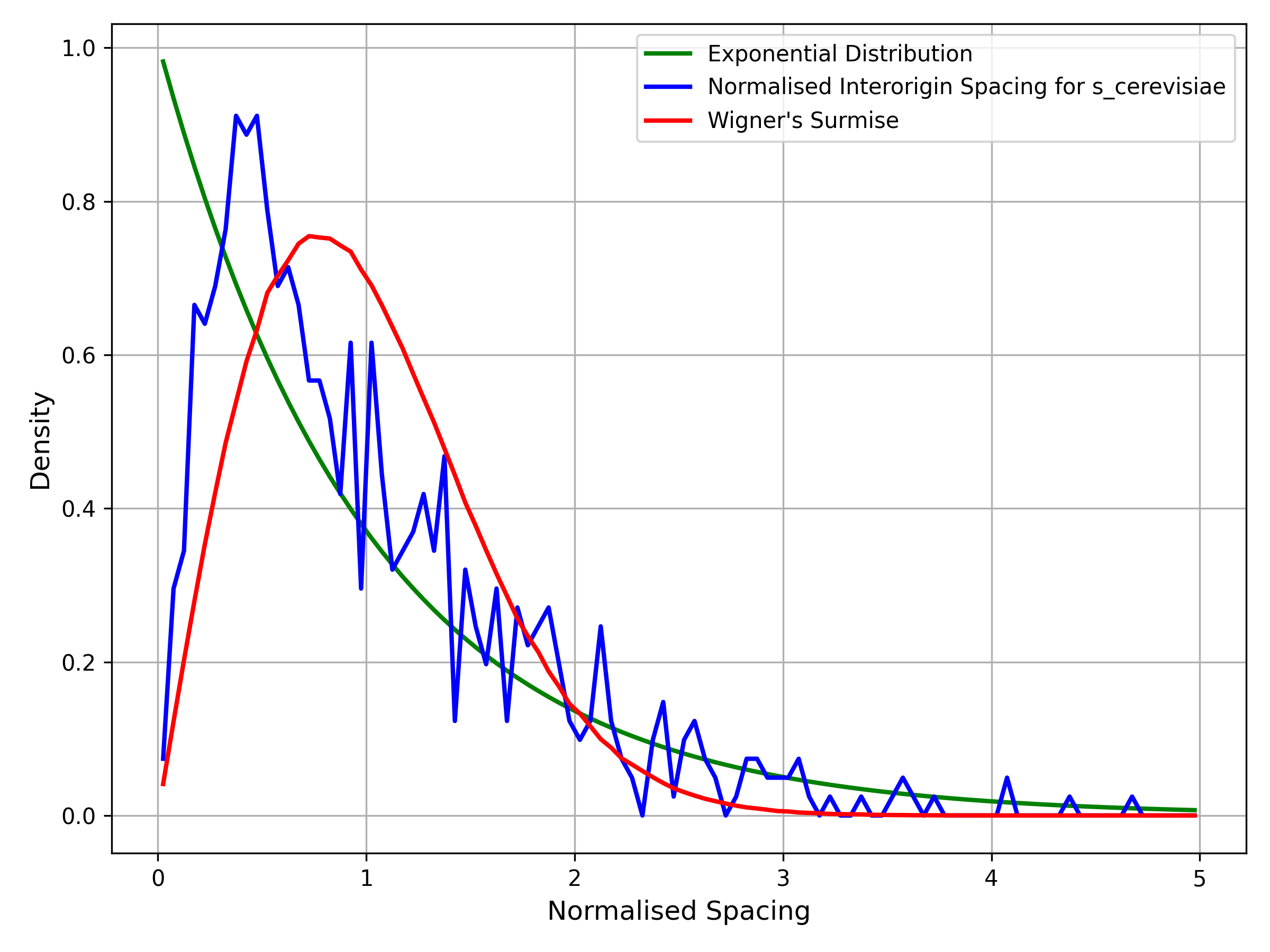} \includegraphics[scale=0.3]{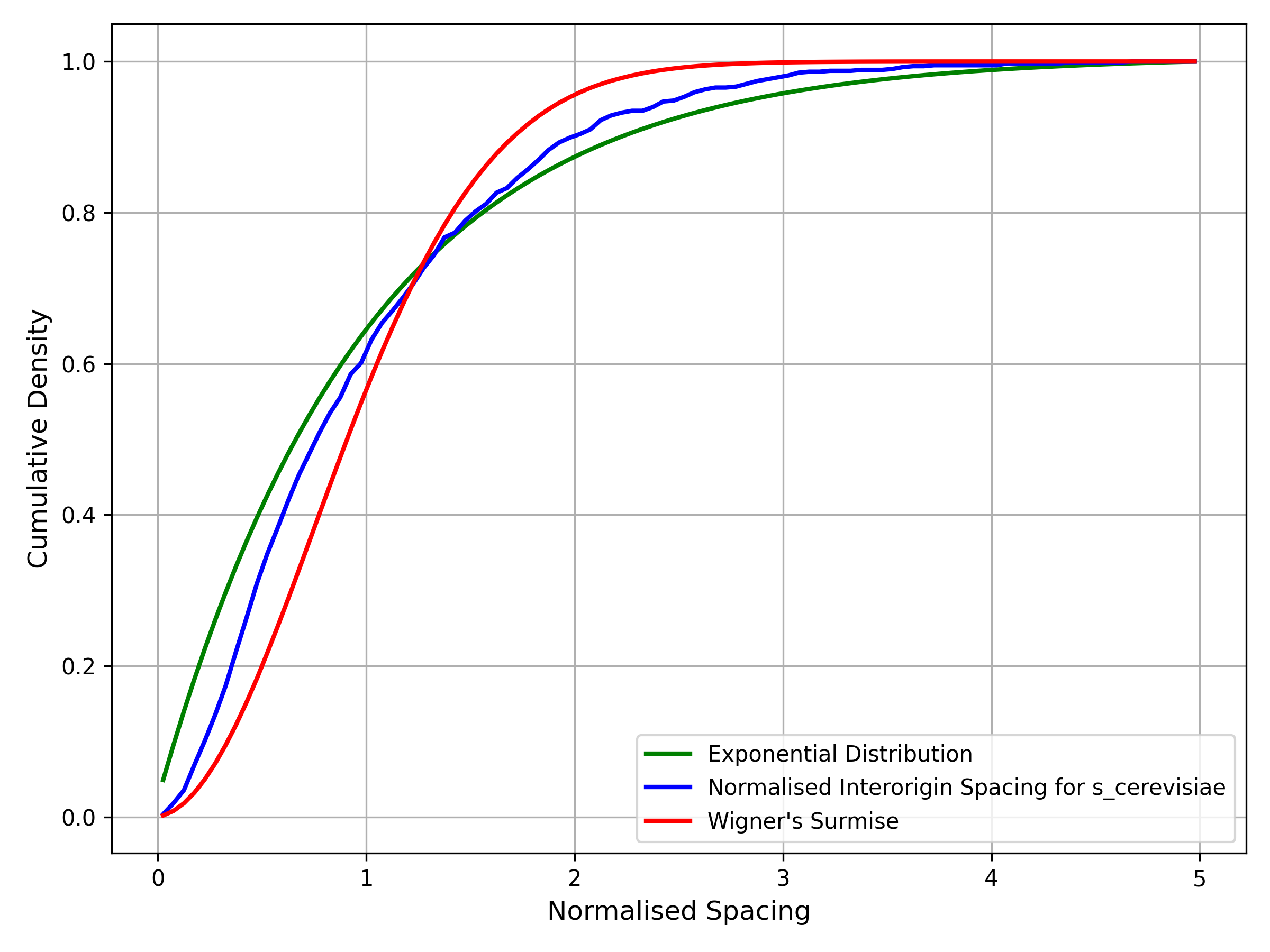}
    \caption{Left: Histogram of re-scaled spacings between midpoints of adjacent replication origins from the yeast strain Saccharomyces cerevisiae (or S cerevisiae), data taken from \cite{Scere}, with Wigner's surmise and exponential distribution for comparison. Right: Cumulative distribution of the same data.  Total Number of Spacings: 813. Number of Chromosomes: 16.}
    \label{fig:scereWig}
\end{figure}

In Figures \ref{fig:KlactisWig}, \ref{fig:lwatiiWig} and \ref{fig:scereWig} we consider  Wigner's surmise, an exponential distribution of parameter $1$, which has mean value 1 (or equivalently; the spacings from a 1D Poisson process of unit intensity)  and spacings between replication origins on chromosomes of certain organisms, re-scaled so that the mean spacing is $1$. This allows for a comparison of the shape of the distributions. We include the cumulative frequency because it gives a smoother curve to work with when the datasets we are working with are small.

In general, fungi and specifically yeast (which are characterised as unicellular fungus) are excellent for intergenetic comparison because there is a lot of variety in species with significant divergence between species (their common ancestor is estimated to be at least 300 million years old) \cite{yeast}. However, they also have shorter chromosomes and generally fewer replication origins, which is problematic as larger data sets allow us to infer with more confidence. 

Kluyveromyces lactis (or K lactis) is a yeast strain often used in industrial applications and genetic studies. We will use origin data taken from \cite{Klactis}. We see the data in Figure \ref{fig:KlactisWig}. 

Lachancea waltii (or L waltii) is another yeast strain and we have used data from \cite{lwaltti2} as shown in Figure \ref{fig:lwatiiWig}.

Figures \ref{fig:KlactisWig} and \ref{fig:lwatiiWig} apparently show a good fit to Wigner's surmise, and this was the observation which started this investigation.  The amount of data for these two species is very limited, however, as can be seen from the unresolved nature of the histogram of the spacing distributions.

Saccharomyces cerevisiae (or S cerevisiae) is one of the most common strains of yeast, commonly referred to as baker's yeast or brewer's yeast, in part due to its role in common types of fermentation. It is extremely well studied in the field of cell biology as evidenced in \cite{Newman} and \cite{lwaltti2} but the data we are using is taken from \cite{Scere} as seen in Figure \ref{fig:scereWig}.  Here we have significantly more data than the previous two organisms. 

Wigner's surmise and the exponential appear not to exactly fit the replication data from S. cerevisiae. However, some sort of interpolation or continuous deformation between Wigner's surmise and the exponential curve might be suitable.  The literature (see the introduction of \cite{kn:musryb15} and references therein) indicates that typically  there are many potential replication origins which are not used, or are so infrequently used in a population that they may not be captured by experimental methods.  This suggests comparison with the so-called ``thinned" random matrix ensembles.  We will consider this model in the next section. 

\section{Comparison with thinned COE statistics}
\label{sect:thinning}

Given a  list of eigenvalues, say \(\{\lambda_{1}, \lambda_{2}, \ldots, \lambda_{N}\}\), we define a thinning parameter \(p\in (0,1]\) and then remove each eigenvalue with probability $1-p$. We can think of this process as having a biased coin toss with probability of heads $p$. We toss the coin for each eigenvalue \(\lambda_{i}\). If the coin lands heads, the eigenvalue stays in our sample. If the coin lands tails, we remove that eigenvalue from our sample.

Thinning of point processes has been studied in general for decades (see \cite{kn:ren56,kn:kal74}) and in random matrix theory (for example \cite{kn:bohpat04,kn:bohpat06,kn:charcla17}). 

Clearly if our thinning parameter $p=1$ then we will not eliminate any of our sample and the nearest neighbour spacing distribution will be approximated by Wigner's surmise. As $p$ gets smaller, we begin eliminating eigenvalues with an increasing likelihood.

Two adjacent eigenvalues (say \(\lambda_{j}\) and \(\lambda_{j+1}\)) from our random matrix sample are correlated - we know they repel linearly (because of the factor of $s$ in Wigner's surmise). If we eliminate an eigenvalue, then the two eigenvalues either side are now neighbours (if we eliminate \(\lambda_{j+1}\) then \(\lambda_{j}\) and \(\lambda_{j+2}\) become neighbours) and their correlation will be weaker than neighbours in the original unthinned sample.

As $p$ approaches $0$, the likelihood of any given eigenvalue being eliminated increases. In practical terms, the thinned sample starts to have increasingly large gaps between eigenvalues. The thinned sample might look like: \(\{\lambda_{1}, \lambda_{50}, \lambda_{94}, \ldots\}\). The bigger the gap between adjacent eigenvalues, the smaller the correlation. If we set $p=0$ then we eliminate the entire sample and get no statistics. However, in the limit as $p$ approaches $0$, the spacing of the thinned sample tends to exponential/Poissonian statistics. This is because the exponential points are not correlated and the process of thinning itself introduces new randomness.

There is no explicit form for the nearest neighbour spacing distribution of a thinned COE ensemble - not even an approximation like Wigner's surmise. Computationally, this process is relatively expensive to implement. One could, for example, generate $200 \times 200$ COE matrices and extract their eigenvalues. The decision has to be made to keep or reject each eigenvalue, and then data from around 10000 matrices could be combined to generate a reasonably smooth approximation to thinned COE eigenvalue spacings.

An alternative to this, which is more effective for small values of $p$, is Bournemann's work from \cite{Fred1} and \cite{Fred2}.

Bournemann's code generates the curve for the probability density function for the spacings between thinned COE eigenvalues, which can be expressed in terms of Fredholm determinants.  

In practical terms, varying the value of $p$ will continuously deform Wigner's surmise into an exponential distribution. We can use this deformation to best fit a thinned plot to  DNA data by finding the optimal value of $p$.

\begin{figure}[H]
    \centering
    \includegraphics[scale=0.35]{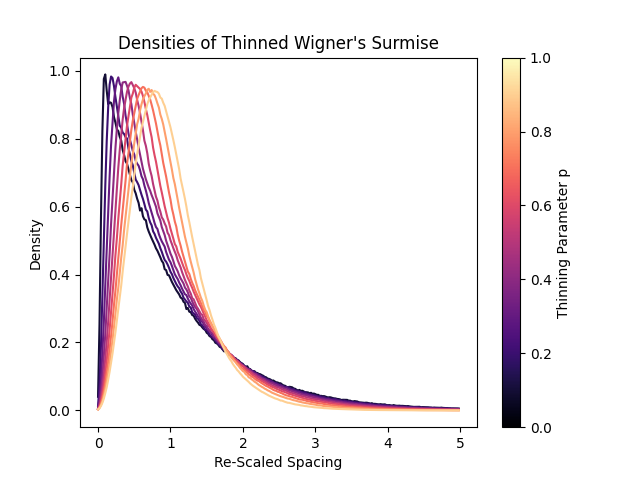} \includegraphics[scale=0.35]{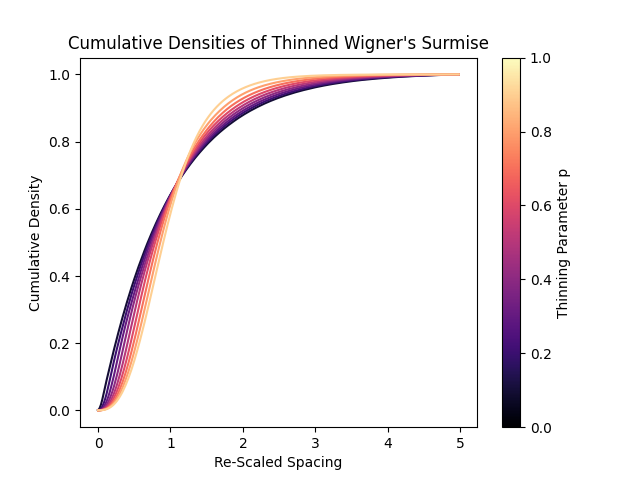}
    \caption{Left: A variety of thinned COE eigenvalue nearest neighbour spacings densities with different thinning parameters, re-scaled to have mean spacing $1$. Lighter yellow indicates close to unthinned ($p \approx 1)$, red/purple indicates moderately thinned ($0.3<p<0.7$) and black indicates very thinned, essentially exponential/Poissonian spacings ($p \approx 0)$.
    The plots are for thinning parameters in a range $0<p<1$ with increments of $0.1$. Right: Corresponding cumulative distributions. }
    \label{fig:denthinscale}
\end{figure}

We know that we can calculate the RMSE between a given genetics dataset and a given distribution, say for example a thinned COE eigenvalue spacing distribution with thinning parameter $p$. It is quite natural to ask what the optimal value of $p$ is to minimise the RMSE between a particular genetics dataset and the thinned COE eigenvalue spacing distribution. For each dataset, we seek to find the optimal thinning parameter to two decimal places. These values and their RMSE are shown for each dataset in Table \ref{tab:ThinOpt}.

\begin{table}[H]
\centering
\begin{tabular}{|c|c|c|}
 \hline
 \multicolumn{3}{|c|}{Optimal Thinning Parameter for Cumulative Density Plots} \\
 \hline
\hline
Sample & Optimal Parameter $p$ (2 d.p) & RMSE (3 s.f)\\
\hline
K lactis & 1.00 & 0.016  \\
\hline
L waltii & 0.95 & 0.011 \\
\hline
S cerevisiae & 0.43 & 0.006\\
\hline
S pombe & 0.28 & 0.013\\
\hline
Drosophila KC & 0.46 & 0.016 \\
\hline
Drosophila S2 & 0.11 & 0.038 \\
\hline 
Mouse ES1 & 0.17 & 0.021  \\
\hline 
Mouse MEF & 0.17 & 0.029\\
\hline
Mouse P19 & 0.21 & 0.035 \\
\hline 
Arabidopsis & 0.22 & 0.026  \\
\hline
Candida CBS138 & 0.03 & 0.009  \\
\hline
Human K562 & 0.08 & 0.189  \\
\hline
Human MCF7 & 0.08 & 0.200 \\
\hline
\end{tabular}
\caption{A table showing for each DNA dataset the optimal thinning parameter for a thinned COE eigenvalue spacing distribution to 2 decimal places and the RMSE between the optimal and the dataset.}
\label{tab:ThinOpt}
\end{table}

\begin{figure}[H]
    \centering
    \includegraphics[scale=0.59]{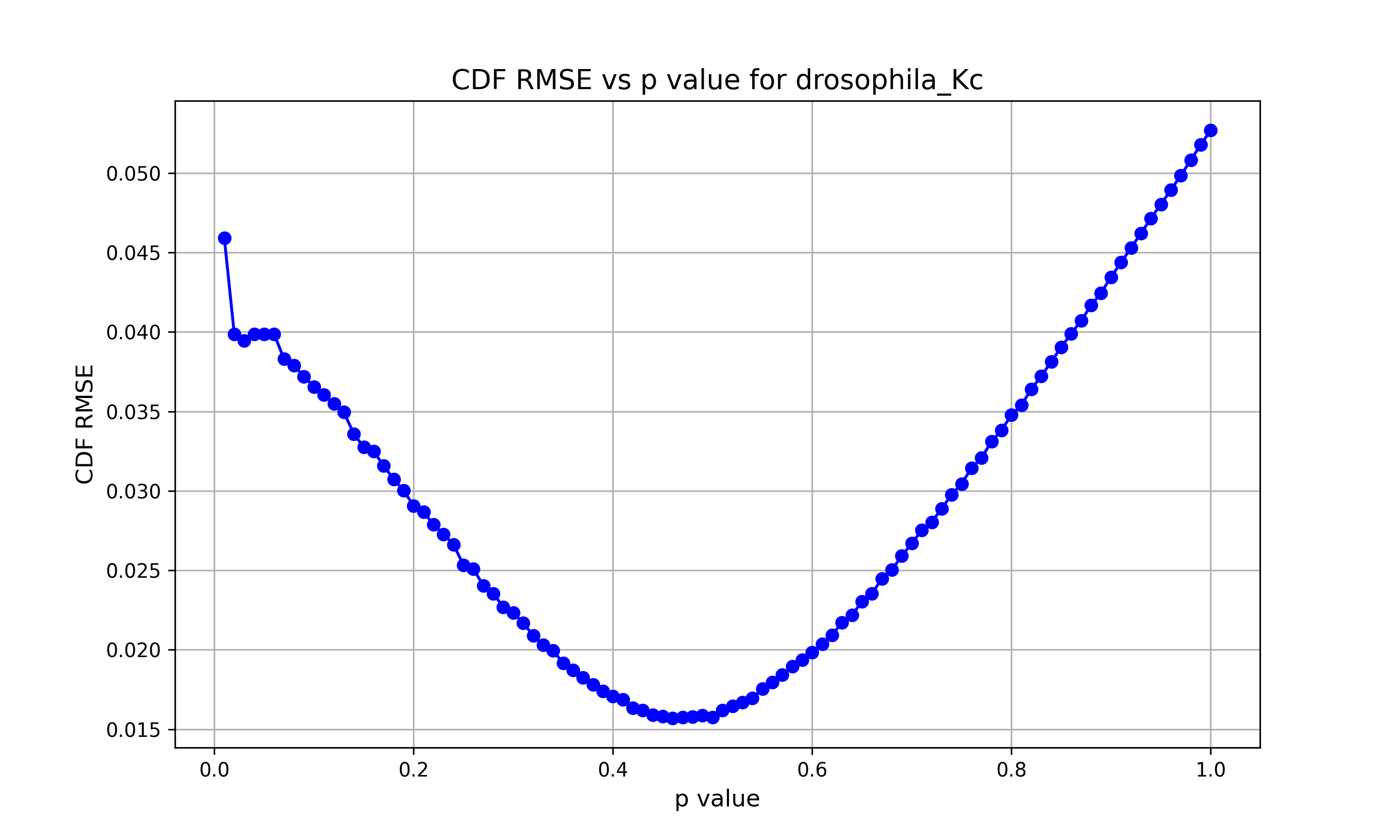}
    \caption{Error profile plotting RMSE between the fruitfly Drosophila KC and a thinned COE eigenvalue spacing distribution for various thinning parameters $0<p<1$. Optimal thinning parameter $p=0.46$ with RMSE $=0.016$.}
    \label{fig:KlactisThinError}
\end{figure}

For each dataset we can produce an error profile showing how the RMSE varies for different thinning parameters, as illustrated in Figure \ref{fig:KlactisThinError}. These give us confidence that a global minimum is the best fit. If, in contrast, this plot was quite erratic with lots of troughs and peaks then it might be less compelling that there was good premise of using these models. Error profiles for the other datasets behave similarly and can be found in the Appendix B of \cite{kn:Day}.

\begin{figure}[H]
    \centering
    \includegraphics[scale=0.33]{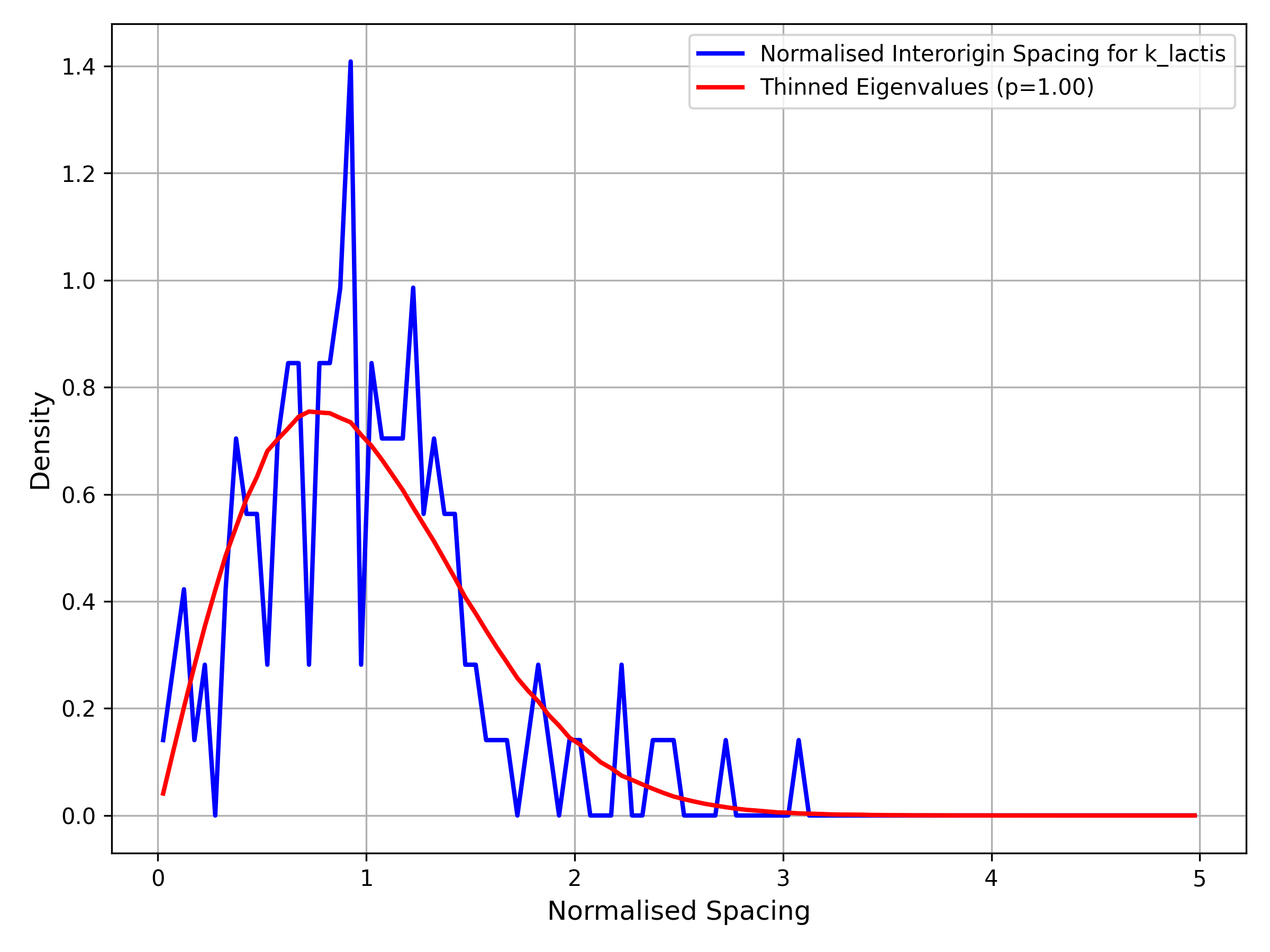} \includegraphics[scale=0.33]{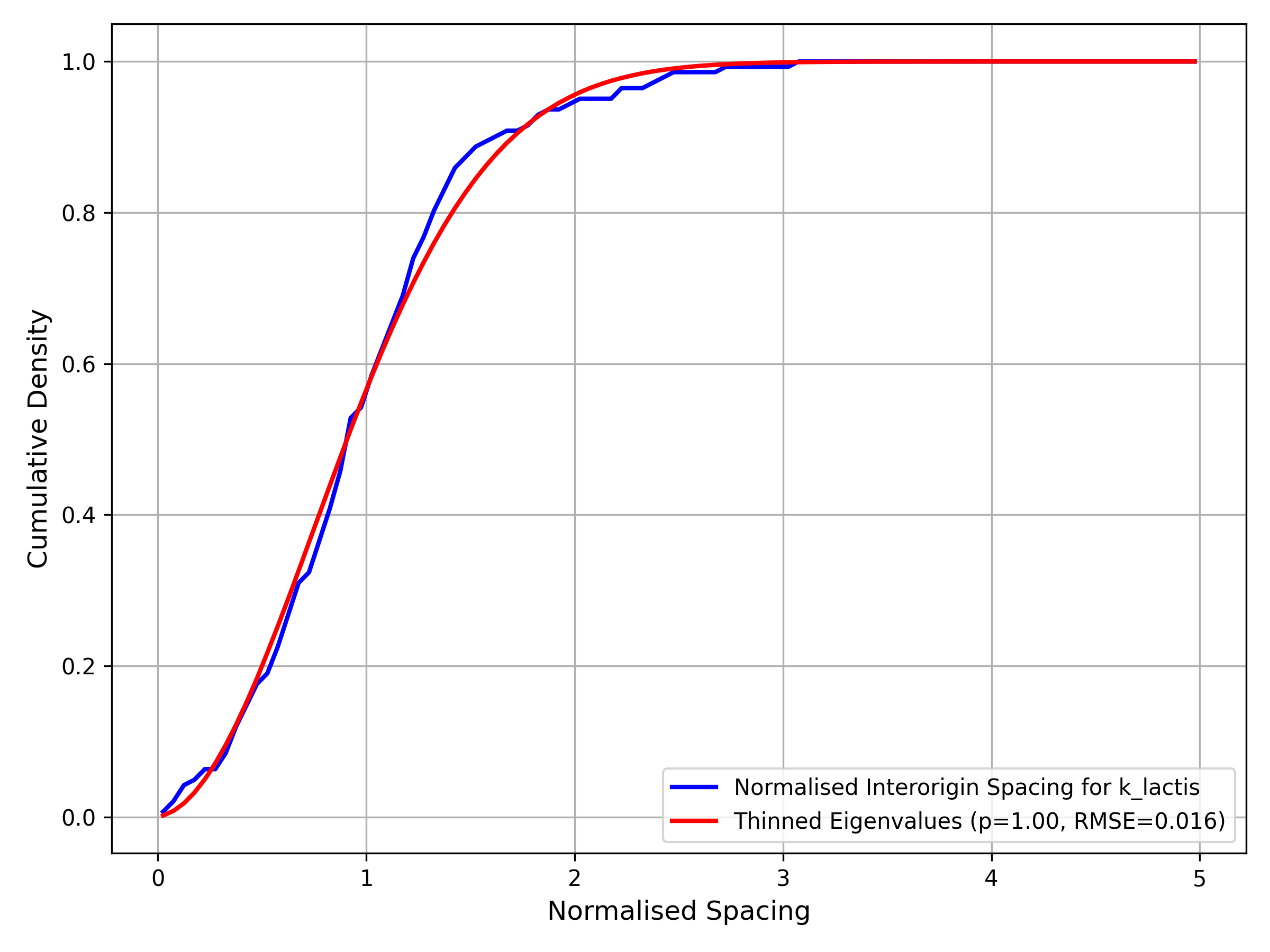}
    \caption{Left: Histogram of re-scaled spacings between midpoints of adjacent replication origins from the yeast strain K Lactis, data taken from \cite{Klactis}, versus thinned COE eigenvalues with parameter $p=1.00$. Right: Corresponding cumulative distributions with RMSE of 0.016.  Total Number of Spacings: 142. Number of Chromosomes: 6.}
    \label{fig:KlactisBestThin}
\end{figure}

\begin{figure}[H]
    \centering
    \includegraphics[scale=0.33]{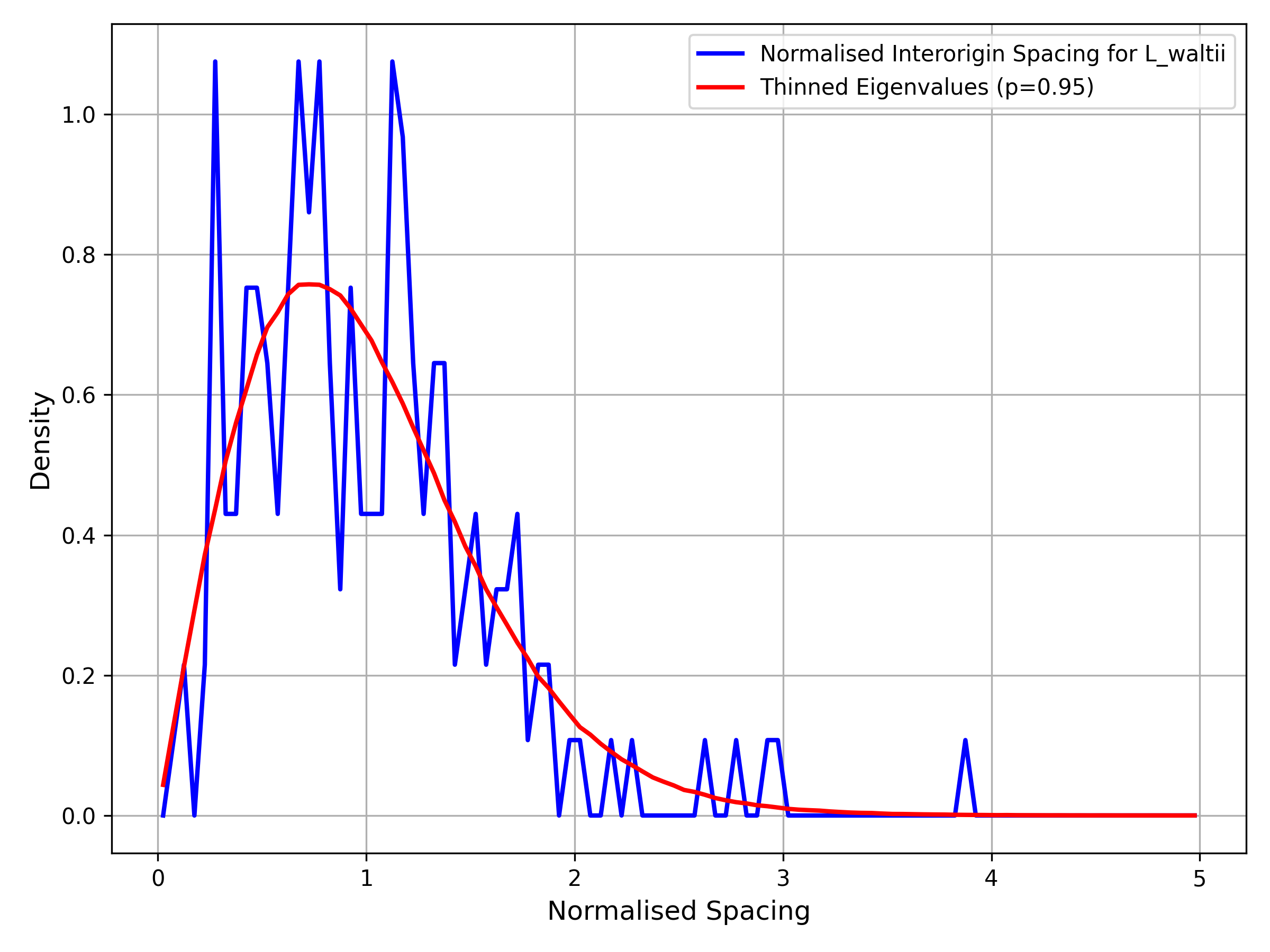} \includegraphics[scale=0.33]{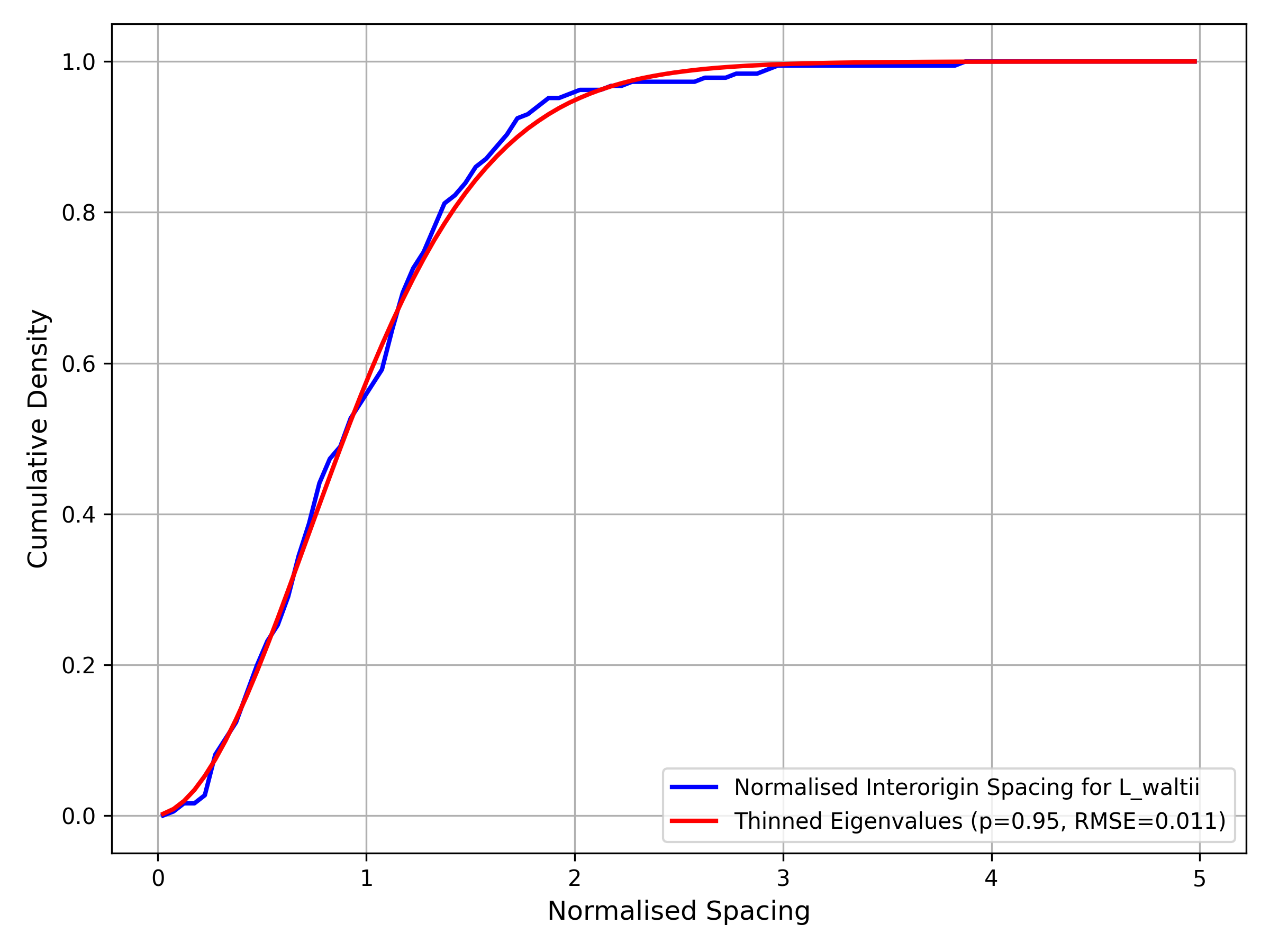}
    \caption{Left: Histogram of re-scaled spacings between midpoints of adjacent replication origins from the yeast strain L Waltii, data taken from \cite{lwaltti2}, versus thinned COE eigenvalues with parameter $p=0.95$. Right: Corresponding cumulative distributions with RMSE of 0.011.  Total Number of Spacings: 186. Number of Chromosomes: 8.}
    \label{fig:LwaltiiBestThin}
\end{figure}

We see for K lactis and L waltii in Figures \ref{fig:KlactisBestThin} and \ref{fig:LwaltiiBestThin} that optimal thinning parameters are $p=1.00$ and $p=0.95$ respectively, meaning that the statistics are quite close to those of the COE without much thinning. The RMSE in both cases is low but we need to exercise caution in drawing too many conclusions since both datasets are so small.

These next datasets are significantly less sparse and allow us to draw conclusions on the statistical distribution of replication origin spacings with more confidence. They all feature so-called model organisms. Model organisms are organisms that are non-human and usually simpler in structure and easier to study than humans \cite{modelorganisms}.

\begin{figure}[H]
    \centering
    \includegraphics[scale=0.33]{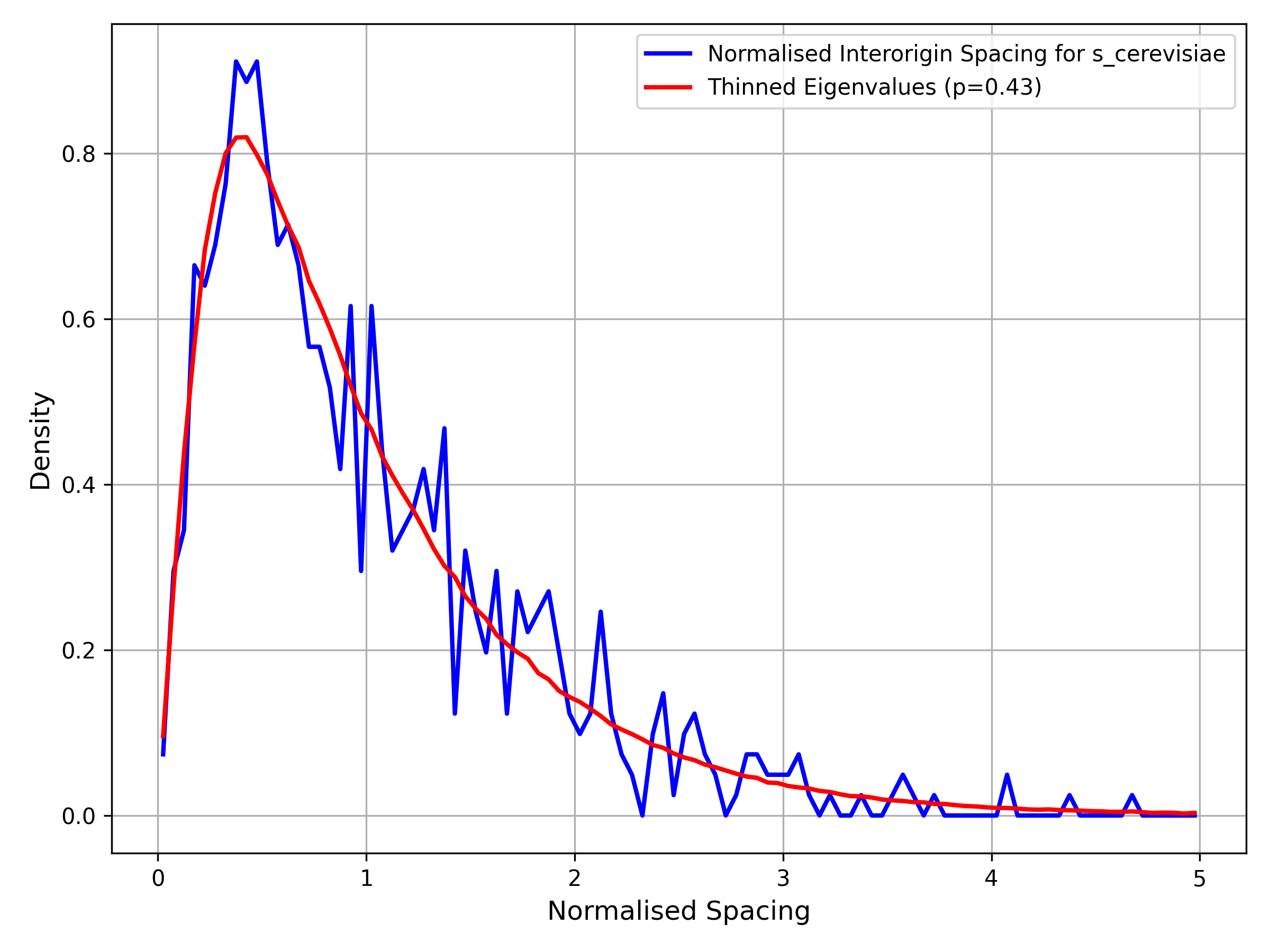} \includegraphics[scale=0.33]{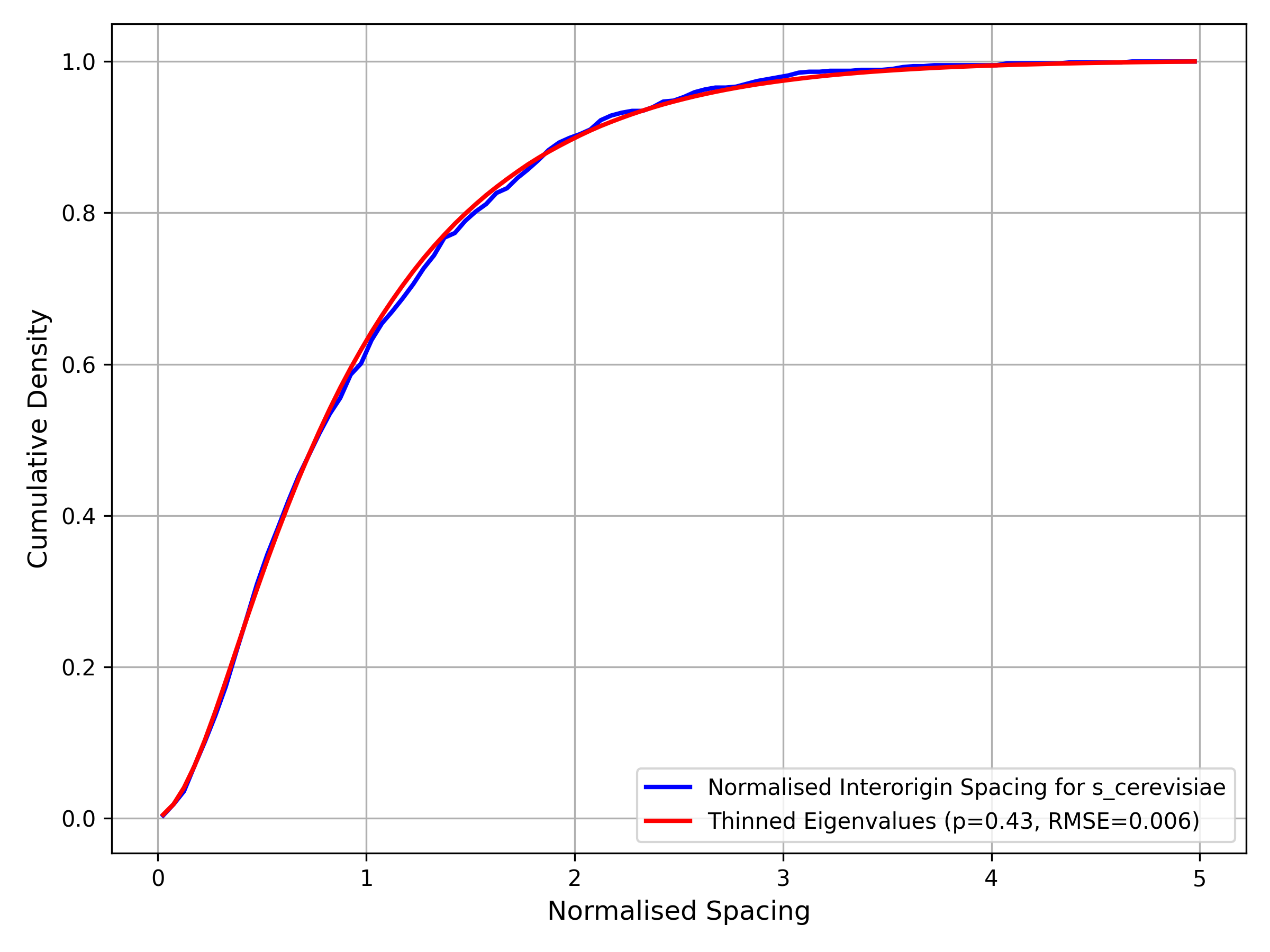}
    \caption{Left: Histogram of re-scaled spacings between midpoints of adjacent replication origins from the yeast strain Saccharomyces cerevisiae (or S cerevisiae), data taken from \cite{Scere}, versus thinned COE eigenvalues with parameter $p=0.43$. Right: Corresponding cumulative distributions with RMSE of 0.006. Total Number of Spacings: 813. Number of Chromosomes: 16.}
    \label{fig:ScereBestThin}
\end{figure}

We have already encountered the yeast S. Cerevisiae in Section \ref{sect:COE}. In Figure \ref{fig:ScereBestThin} we see the replication data plotted against the thinned COE with the best fit. We see that the thinning parameter that produces the best fit is $p=0.43$, which is much more significant thinning than either of the previous yeast varieties. In Figure \ref{fig:scereWig} we saw that the S. Cerevisiae data was not a good fit to Wigner's surmise, but that thinning with $p=0.43$ gives a much better fit. 

\begin{figure}[H]
    \centering
    \includegraphics[scale=0.33]{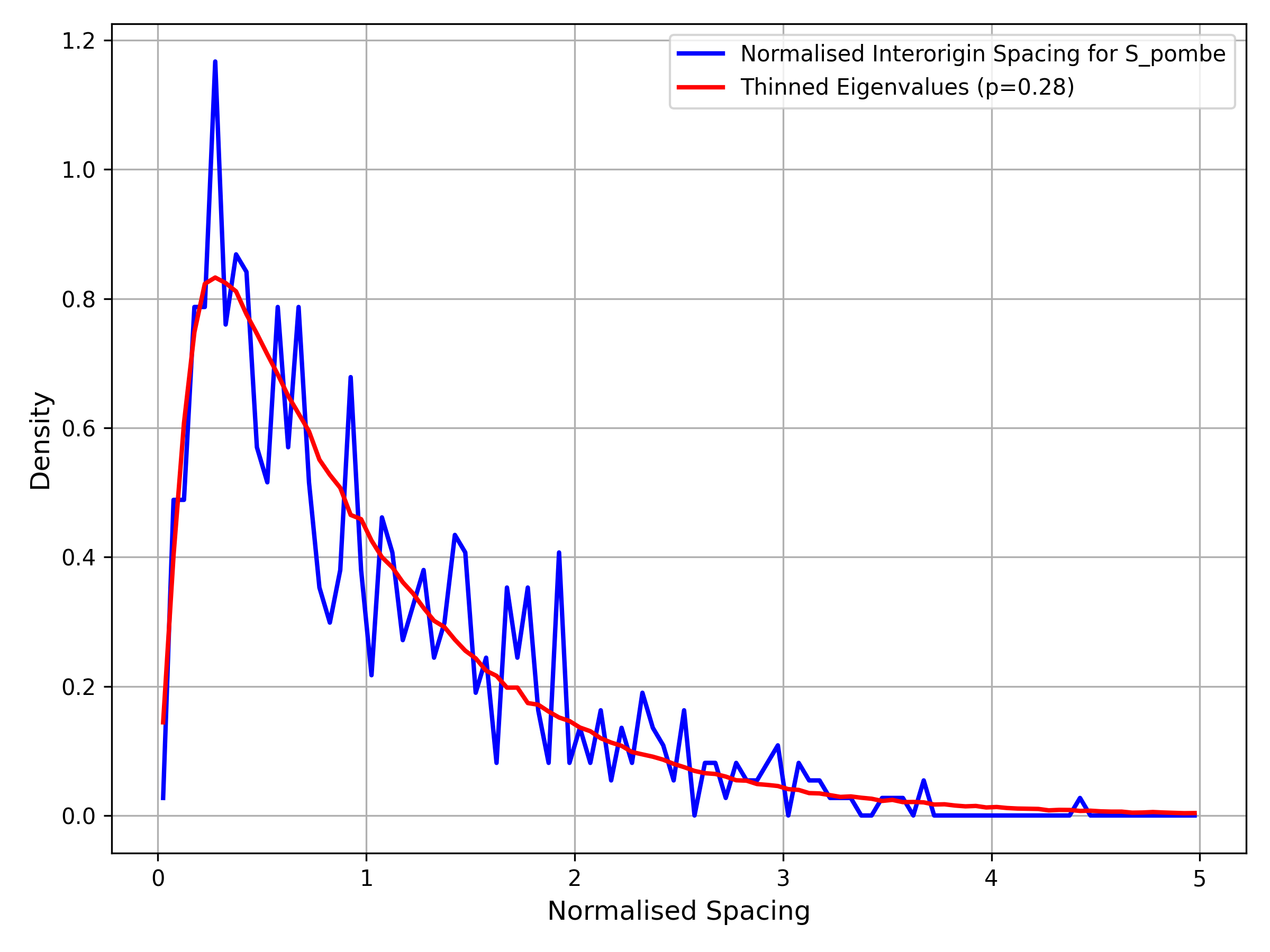} \includegraphics[scale=0.33]{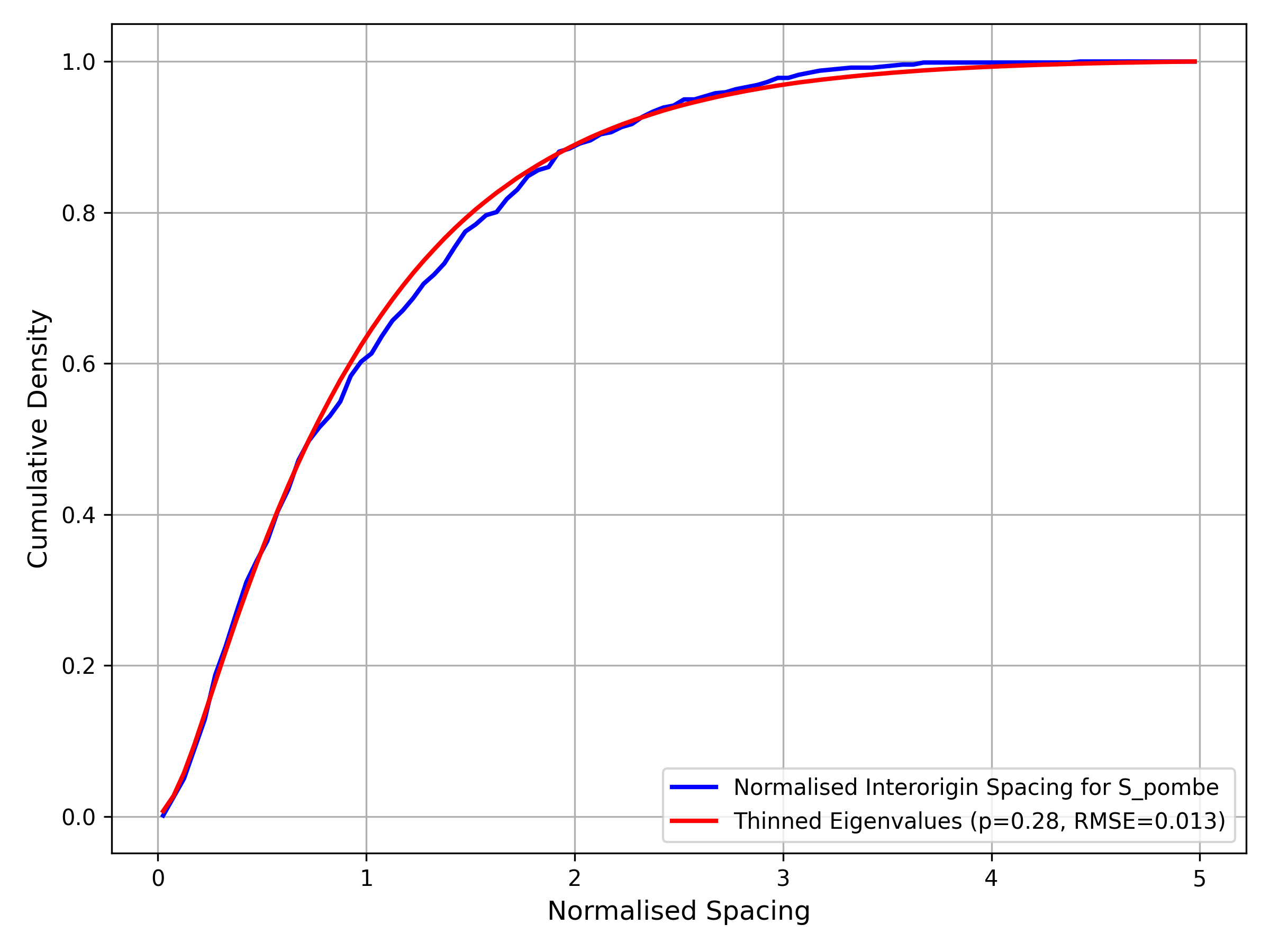}
    \caption{Left: Histogram of re-scaled spacings between midpoints of adjacent replication origins from the yeast strain Schizosaccharomyces pombe (or S pombe), data taken from \cite{Scere}, versus thinned COE eigenvalues with parameter $p=0.28$. Right: Corresponding cumulative distributions with RMSE of 0.013. Total Number of Spacings: 738. Number of Chromosomes: 3.}
    \label{fig:SpombeBestThin}
\end{figure}

Another yeast for which we have data is Schizosaccharomyces pombe (or S pombe), a fission yeast, and the replication origin distribution is shown in Figure \ref{fig:SpombeBestThin}.  

\begin{figure}[H]
    \centering
    \includegraphics[scale=0.33]{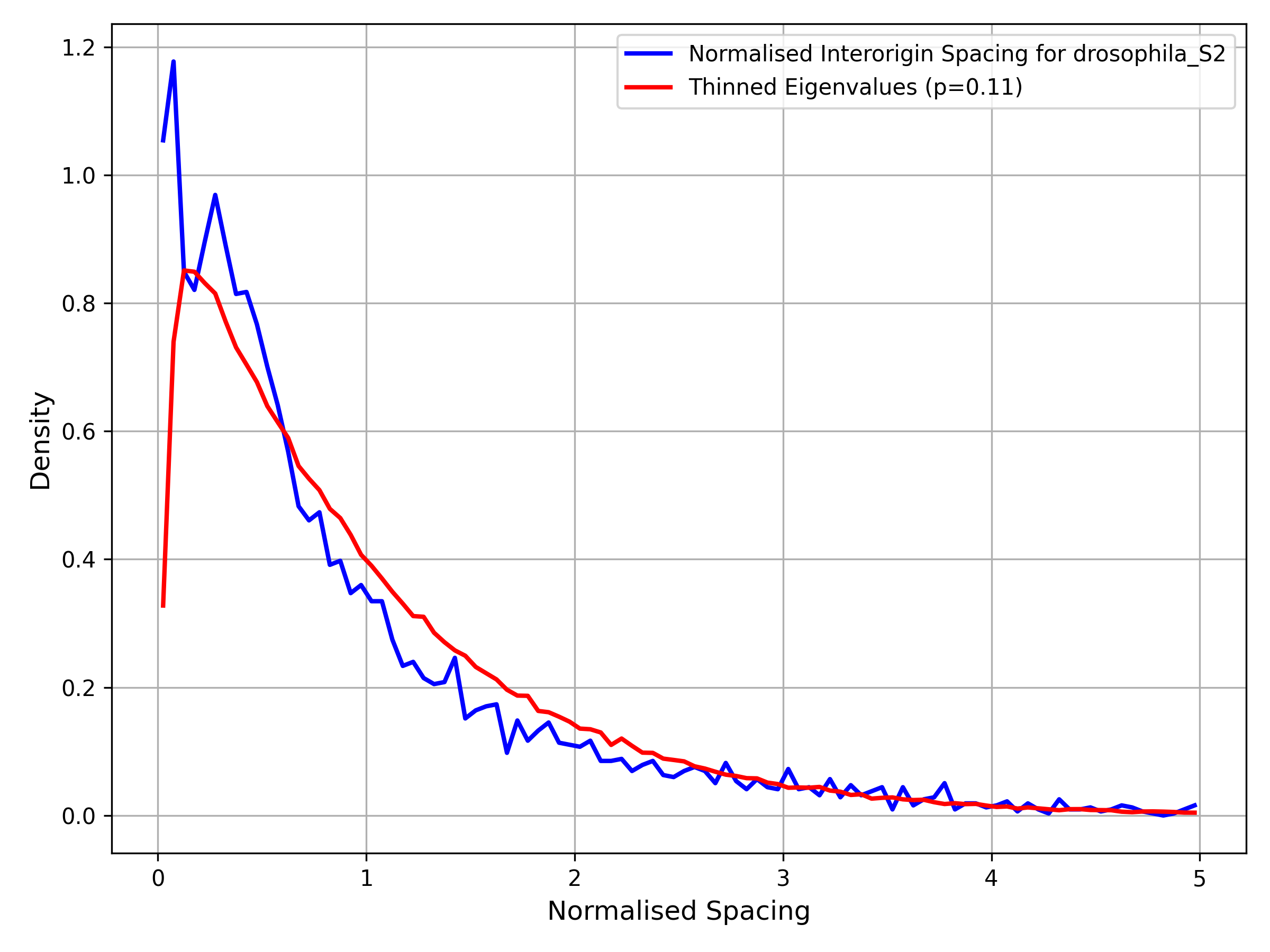} \includegraphics[scale=0.33]{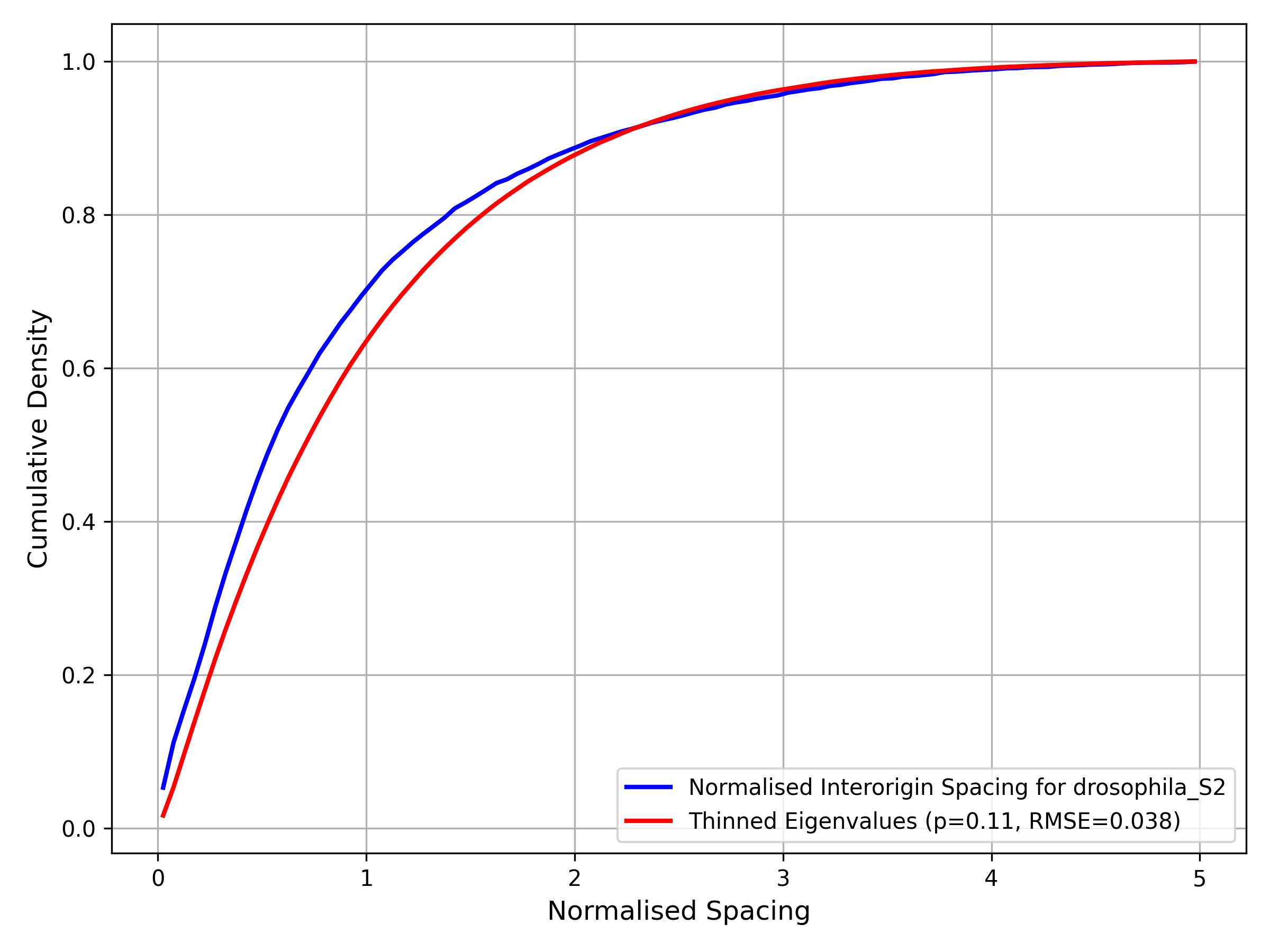}
    \caption{Left: Histogram of re-scaled spacings between midpoints of adjacent replication origins from the fruit fly Drosophila melanogaster cell line Schneider 2 (or Drosophila S2), data taken from \cite{Drosophila}, versus thinned COE eigenvalues with parameter $p=0.11$. Right: Corresponding cumulative distributions with RMSE of 0.038. Total Number of Spacings: 6450. Number of Chromosomes: 6.}
    \label{fig:DrosoS2BestThin}
\end{figure}

\begin{figure}[H]
    \centering
    \includegraphics[scale=0.33]{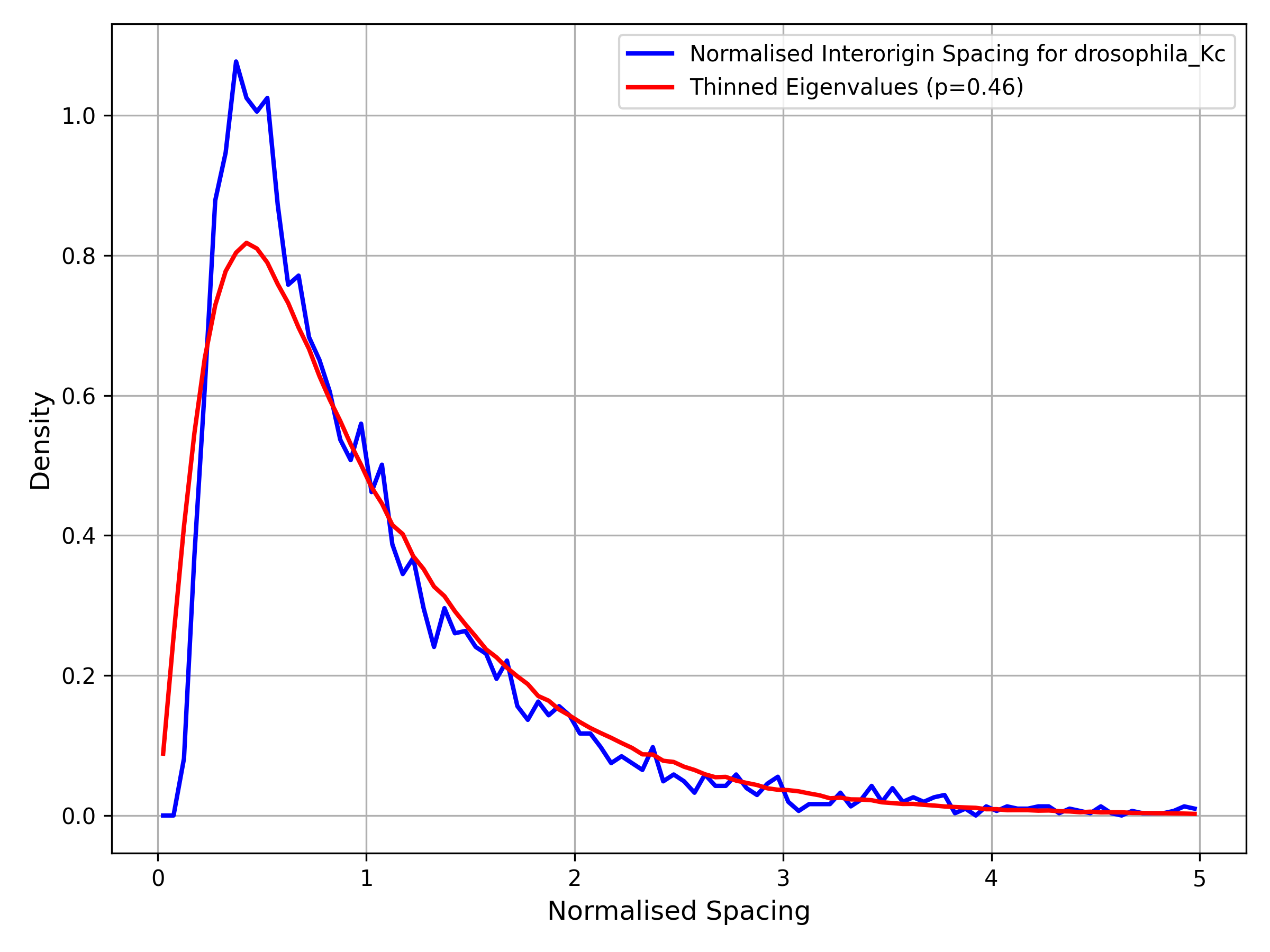} \includegraphics[scale=0.33]{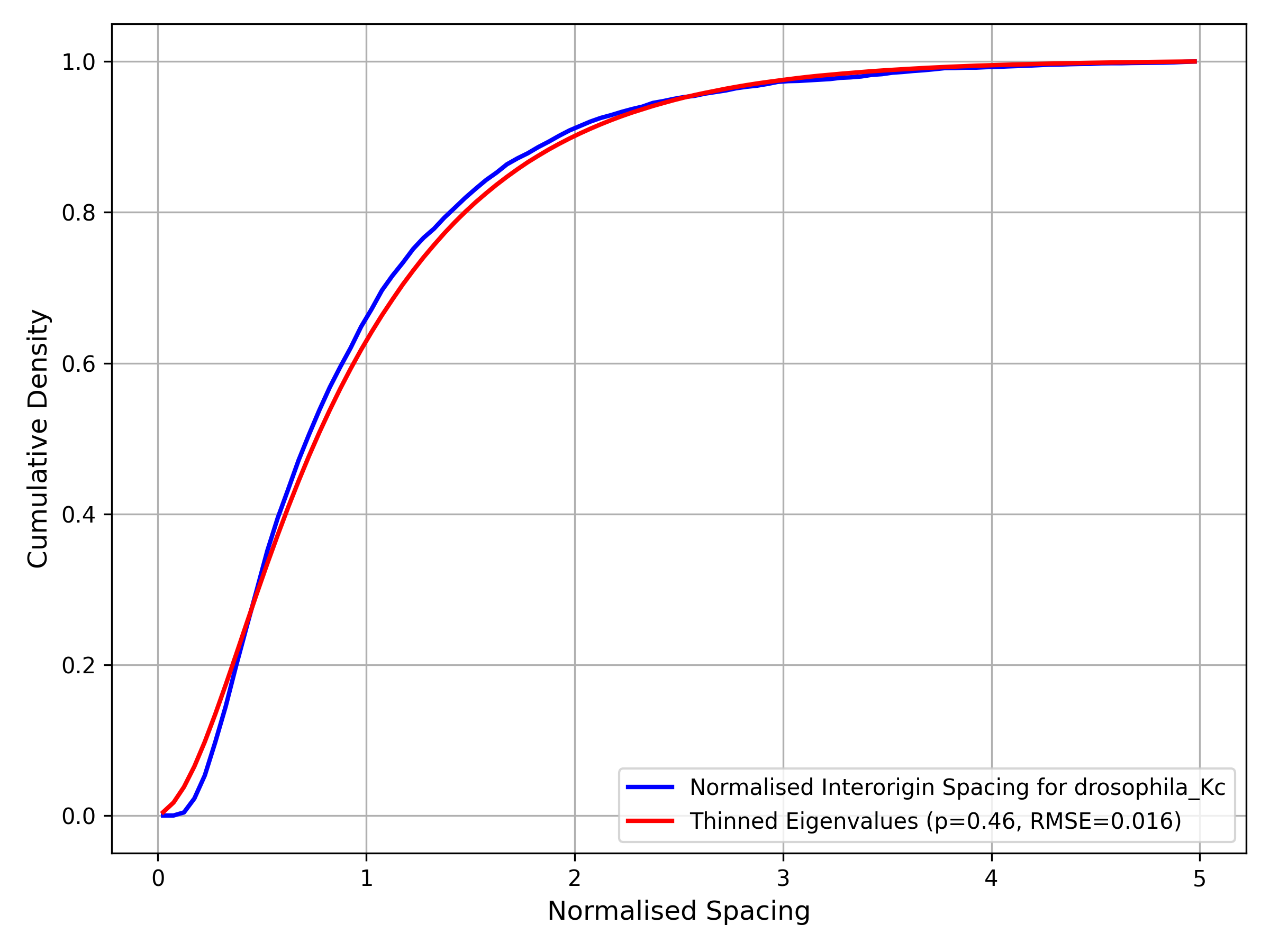}
    \caption{Left: Histogram of re-scaled spacings between midpoints of adjacent replication origins from the fruit fly Drosophila melanogaster cell line KC167 (or Drosophila KC), data taken from \cite{Drosophila}, versus thinned COE eigenvalues with parameter $p=0.46$. Right: Corresponding cumulative distributions with RMSE of 0.016. Total Number of Spacings: 6178, Number of Chromosomes: 6.}
    \label{fig:DrosoKCBestThin}
\end{figure}

Drosophila melanogaster (or just Drosophila) is a species of fly, specifically the fruit fly. It is an ideal model organism as it is reasonably simple genetics, a short life cycle and large reproductive capacity (equivalently, one expects a large number of offspring from a single generation) \cite{sang2001drosophila}. Drosophila Schneider 2 cells (more commonly Drosophila S2) and KC167 (more commonly Drosophila KC) are two of the more commonly sampled cell lines and can be seen in Figures \ref{fig:DrosoS2BestThin} and \ref{fig:DrosoKCBestThin}. More detail on their  differences can be found in \cite{drosodifferences}.

For Drosophila S2 in Figure \ref{fig:DrosoS2BestThin}, we don't see evidence of the replication origins avoiding being close together; the best fit is an extreme thinning of COE eigenvalues, very close to the distribution of uncorrelated points.  There does not seem to be a thinning of the COE that produces a good fit for the data.  However, for S cerevisiae, S Pombe and Drosophila KC we see in Figures \ref{fig:ScereBestThin}, \ref{fig:SpombeBestThin} and \ref{fig:DrosoKCBestThin}  some correspondence with thinned models. We have smooth curves,  low RMSEs  and thinning parameters of $p=0.43$,  $p=0.28$ and $p=0.46$ respectively. There is clearly still some correlation between points but these datasets are perhaps the most compelling to suggest using uniform thinning as a viable model.

\begin{figure}[H]
    \centering
    \includegraphics[scale=0.7]{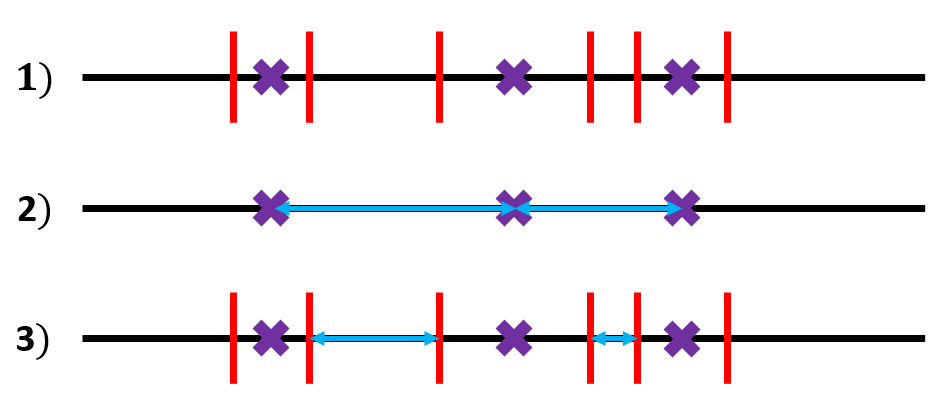}
    \caption{A figure showing 1) replication origins with their ends (red vertical lines) and midpoint (purple crosses) depicted, 2) their mid point to mid point spacing with blue arrows and 3) their end to end spacings with blue arrows.}
    \label{fig:endtoend}
\end{figure}

Recall that we modeled each replication origin as a point by taking the midpoint of its start and end point. This is common in the literature. In \cite{EndtoEnd} there is in depth analysis of an alternative approach, which considers so called end to end spacings. This method considers spacings between neighbouring replication origins as the distance from the right most point of one origin to the left most of the next origin. This is illustrated in Figure \ref{fig:endtoend}. This project focuses on using midpoint spacings because this is how the majority of the genetics literature approaches this problem, but also from a mathematical modeling perspective it is  easier to model each replication origin as a single point with zero width. However, if we are investigating spacings sufficiently small that they are of comparable size of the width of replication origins, it becomes important to understand the repercussions of this choice. 
    
\begin{figure}[H]
        \centering
        \includegraphics[scale=0.5]{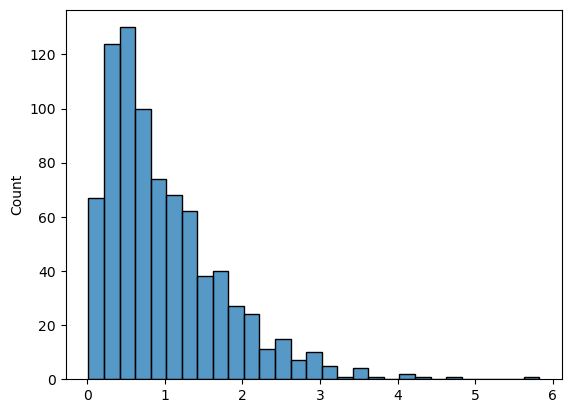}
        \includegraphics[scale=0.5]{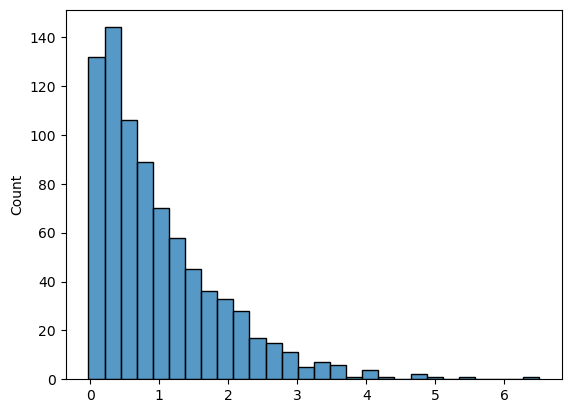}
        \caption{A histogram comparison of S cerevisiae spacing data. The left plot is the midpoint to midpoint distance data, where there are fewer spacings in the left most bin. The right plot is the end to end distance data, where the left most bar is much higher. Both datasets are re-scaled to have mean spacing one.}
        \label{fig:midpointendtoend}
    \end{figure}

    Consider Figure \ref{fig:endtoend} where  origins are close enough that their width is a significant proportion of their spacings.  Choosing to measure midpoint to midpoint versus end to end can make an observable difference to the nearest neighbour spacing, particularly for the  smallest spacings, as in Figure \ref{fig:midpointendtoend}.

\begin{figure}[H]
    \centering    
    \includegraphics{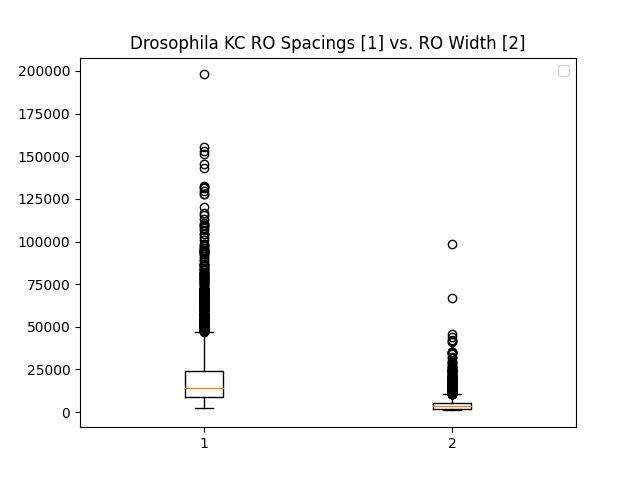}
    \caption{Side-by-side boxplot comparison of the replication origin spacing on the left (i.e. distance between midpoints of neighbouring origins) and replication origin length  from Drosophila KC, on the right. 8.24 \% of the Drosophila KC origin lengths are within the interquartile range of the spacings. The vertical axis is measured in base pairs.}
    \label{fig:drosoKCCompare}
\end{figure}

In Figure \ref{fig:drosoKCCompare} we display a boxplot of Drosophila KC replication origin length next to the boxplot of  replication origin spacings (using distances between midpoints). 8.24 \% of the Drosophila KC origin lengths overlap with the interquartile range of the spacings. In contrast, in the K lactis dataset, there are no origins of length within the interquartile range of the spacings and 0.34 \% of the Human MCF7 origin lengths are within the interquartile range of their respective spacings. For the Drosophila KC dataset, the overlap provides a viable explanation for the lack of small interorigin spacings. Drosophila KC is the only dataset of those we are considering which has such a large overlap between the smallest spacings between the midpoints of neighbouring replication origins and the largest replication origin widths. 

The box plots in Figure \ref{fig:drosoKCCompare} indicate the interquartile range of the data, with the orange line representing the median: so a quarter of the data points lie between the median and the top of the box, with another quarter lying between the median and the bottom of the box. The threshold for outliers is $1.5$  times the interquartile range (the height of the black box) above or below the upper and lower quartiles, and is marked by the black `whiskers'. Values from the dataset that are outliers are marked as black circles.

The mouse is the most common model organism for pre-clinical studies even though it has not proven particularly reliable at predicting the outcome of studies in humans. Mice genomes are extremely similar to the human genome with a $99\%$ overlap. Mice being relatively small allows for large scale studies with a high output making them a cost-efficient organism \cite{mouseuse}. 

All of the data from mice genomes is taken from \cite{Drosophila}. Each of the samples only look at a single chromosome but compensate somewhat with a large number of replication origin spacings.

\begin{figure}[H]
    \centering
    \includegraphics[scale=0.33]{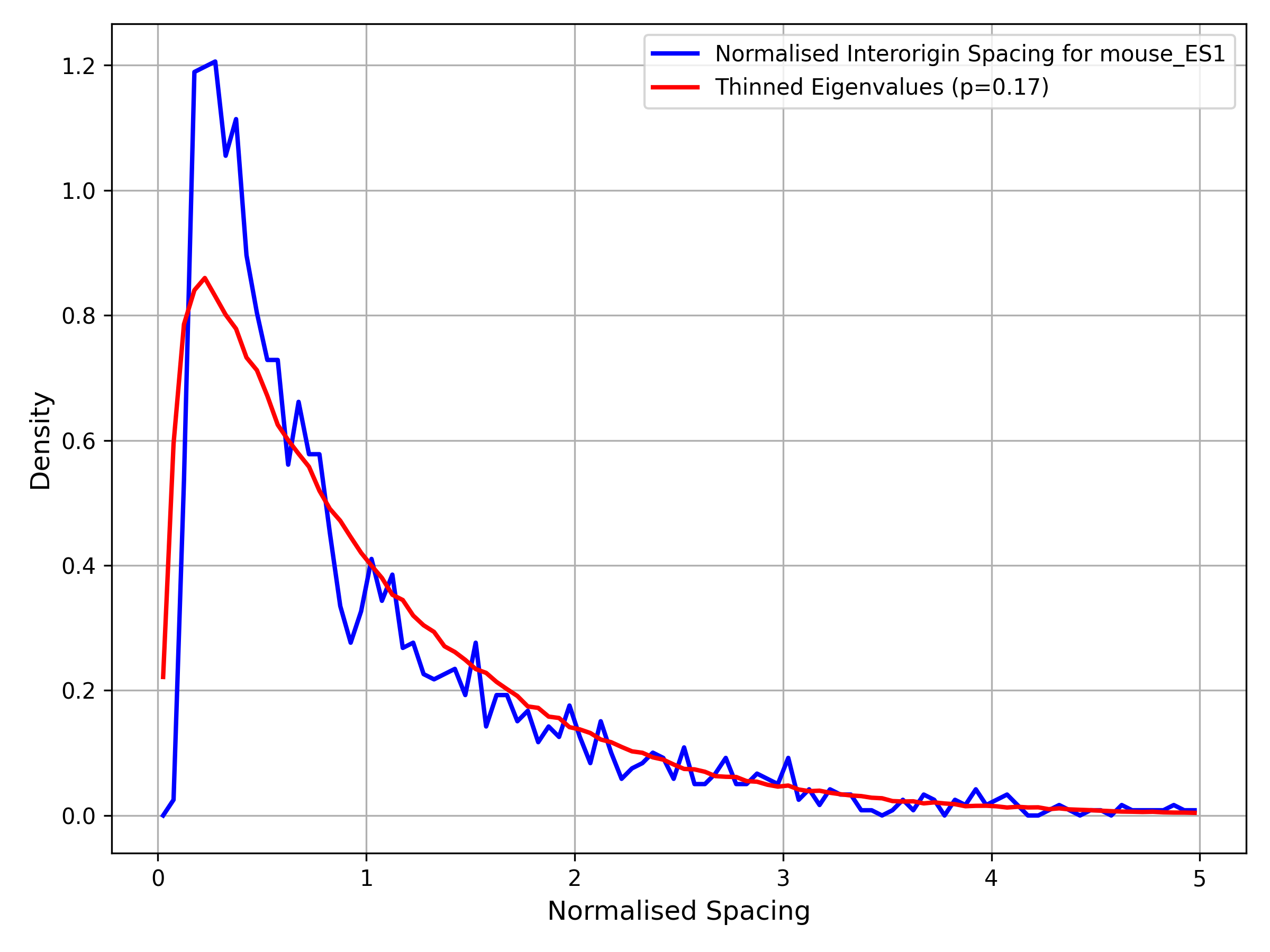} \includegraphics[scale=0.33]{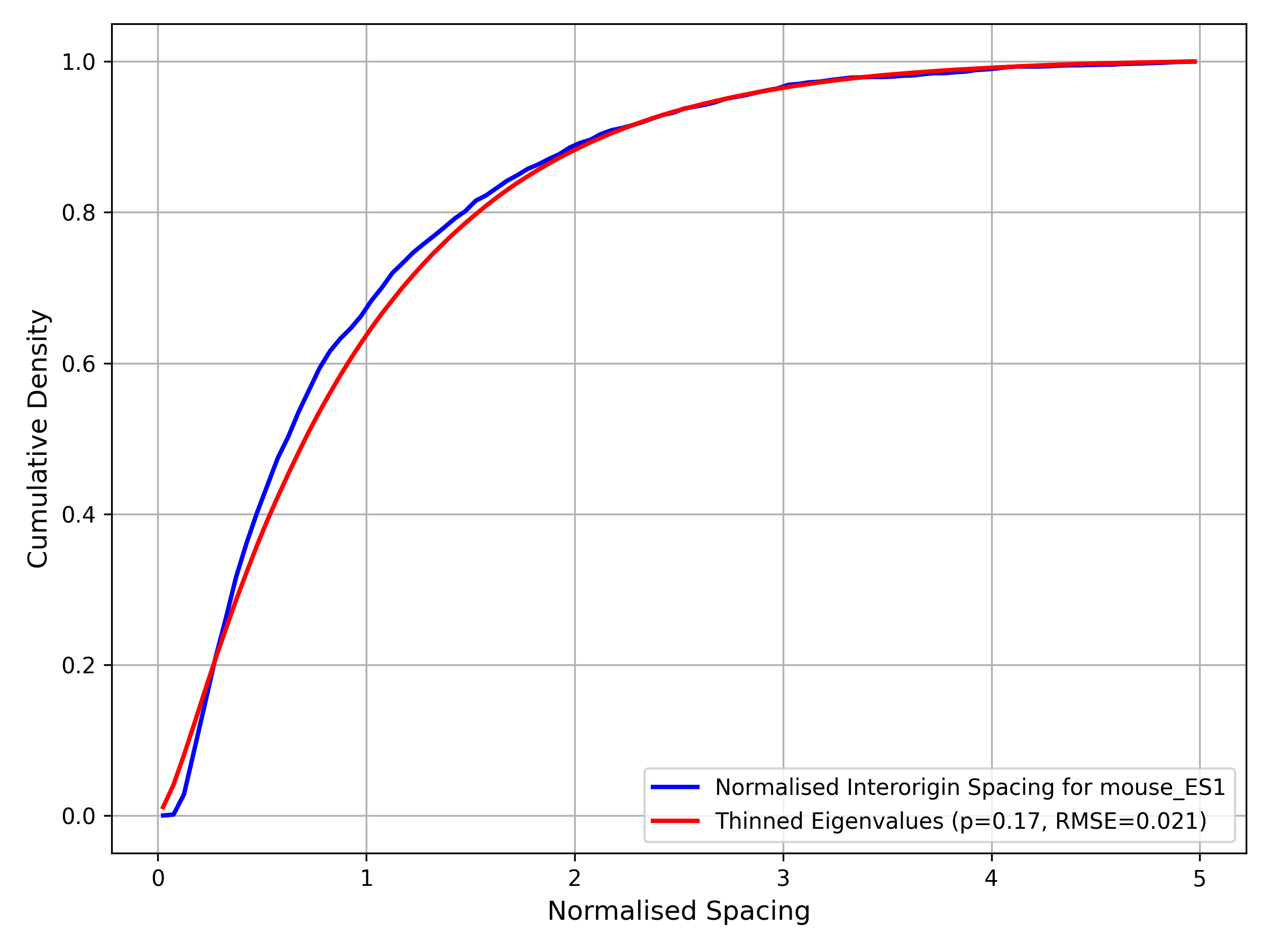}
    \caption{Left: Histogram of re-scaled spacings between midpoints of adjacent replication origins from the mouse embryonic cells (or Mouse ES1), data taken from \cite{Drosophila}, versus thinned COE eigenvalues with parameter $p=0.17$. Right: Corresponding cumulative distributions with RMSE of 0.021. Total Number of Spacings: 2411. Number of Chromosomes: 1.}
    \label{fig:MouseES1BestThin}
\end{figure}

\begin{figure}[H]
    \centering
    \includegraphics[scale=0.33]{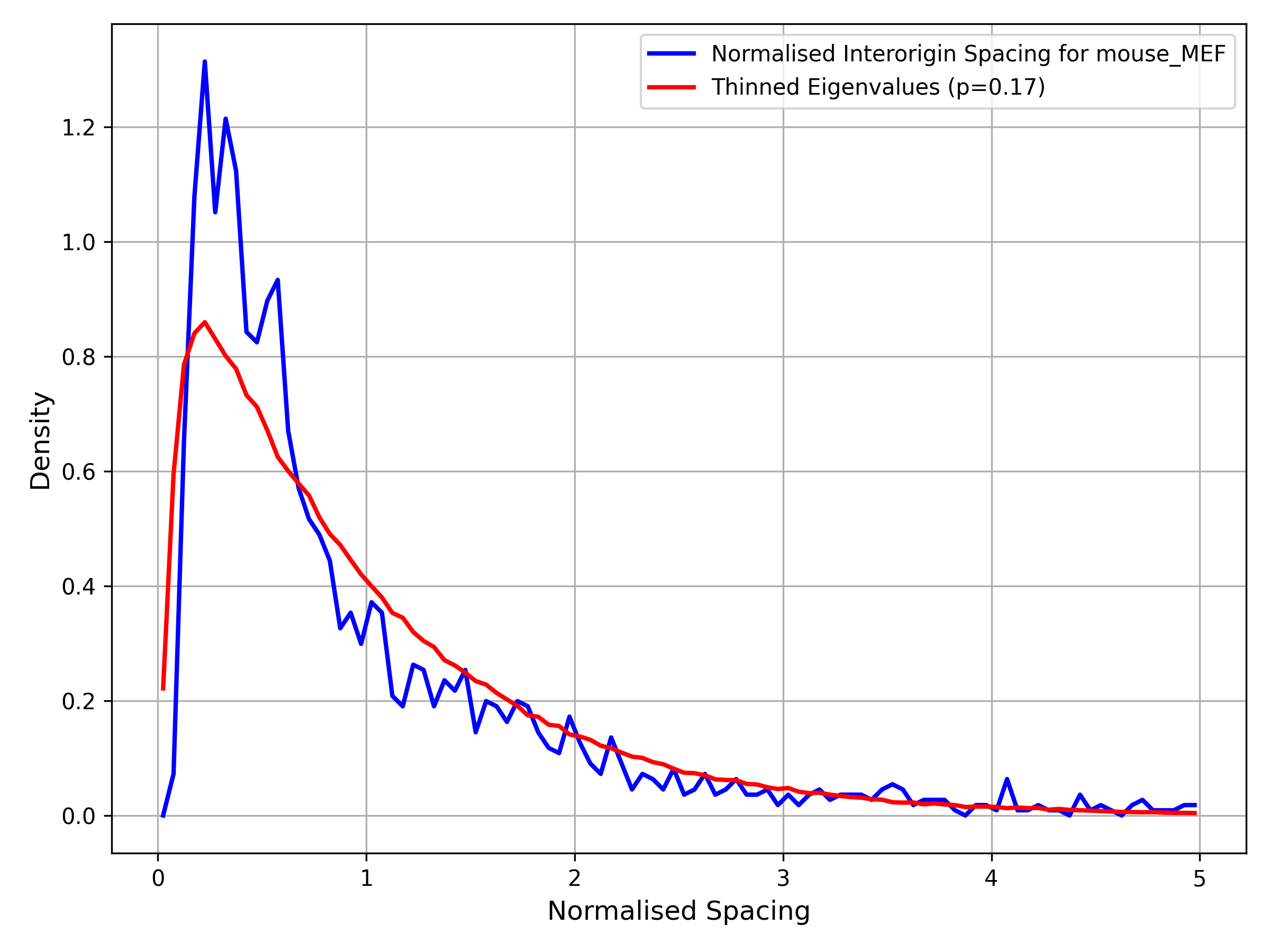} \includegraphics[scale=0.33]{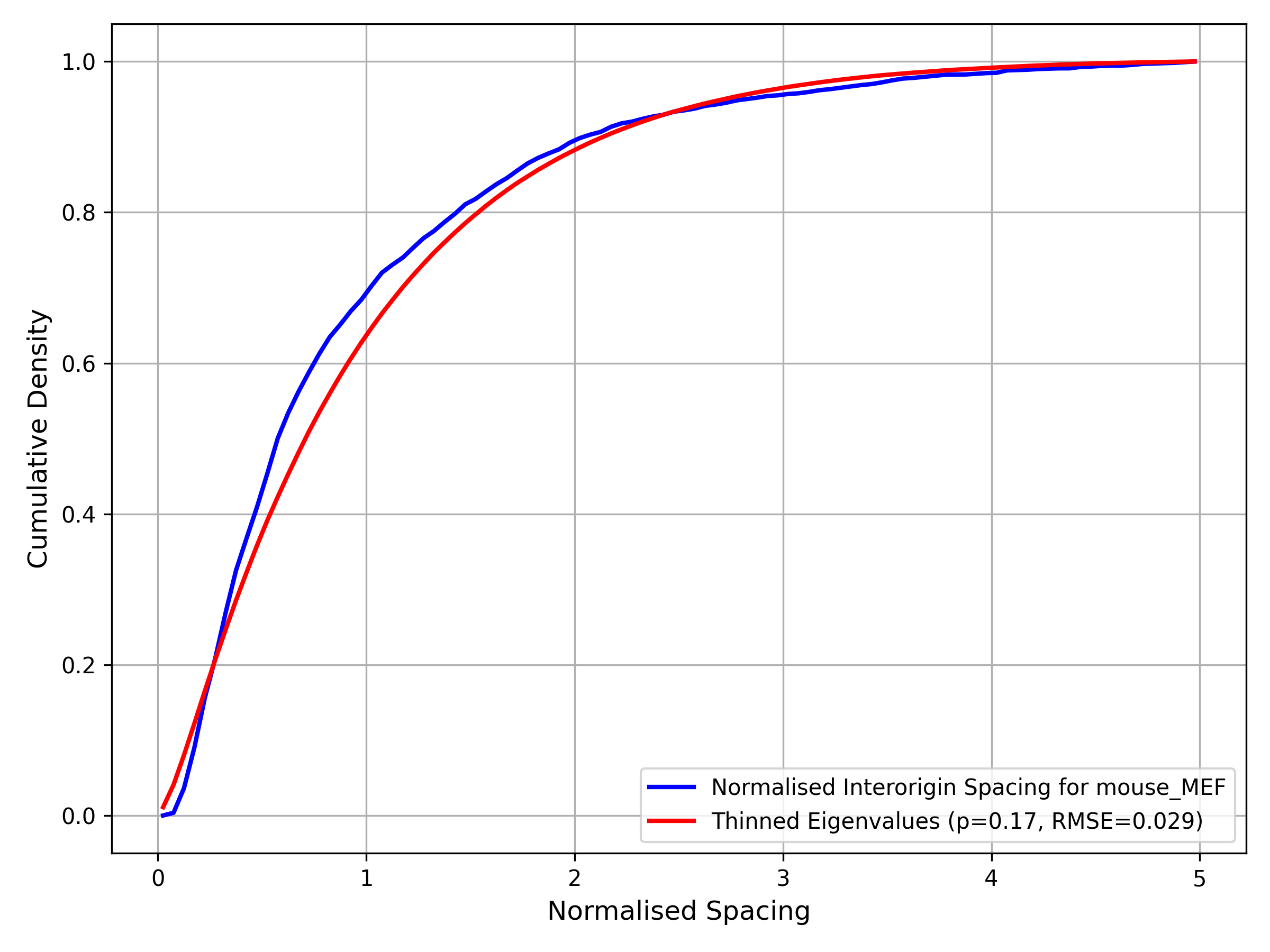}
    \caption{Left: Histogram of re-scaled spacings between midpoints of adjacent replication origins from mouse teratocarcinoma cells (or Mouse MEF), data taken from \cite{Drosophila}, versus thinned COE eigenvalues with parameter $p=0.17$. Right: Corresponding cumulative distributions with RMSE of 0.029. Total Number of Spacings: 2230. Number of Chromosomes: 1.}
    \label{fig:MouseMEFBestThin}
\end{figure}

\begin{figure}[H]
    \centering
    \includegraphics[scale=0.33]{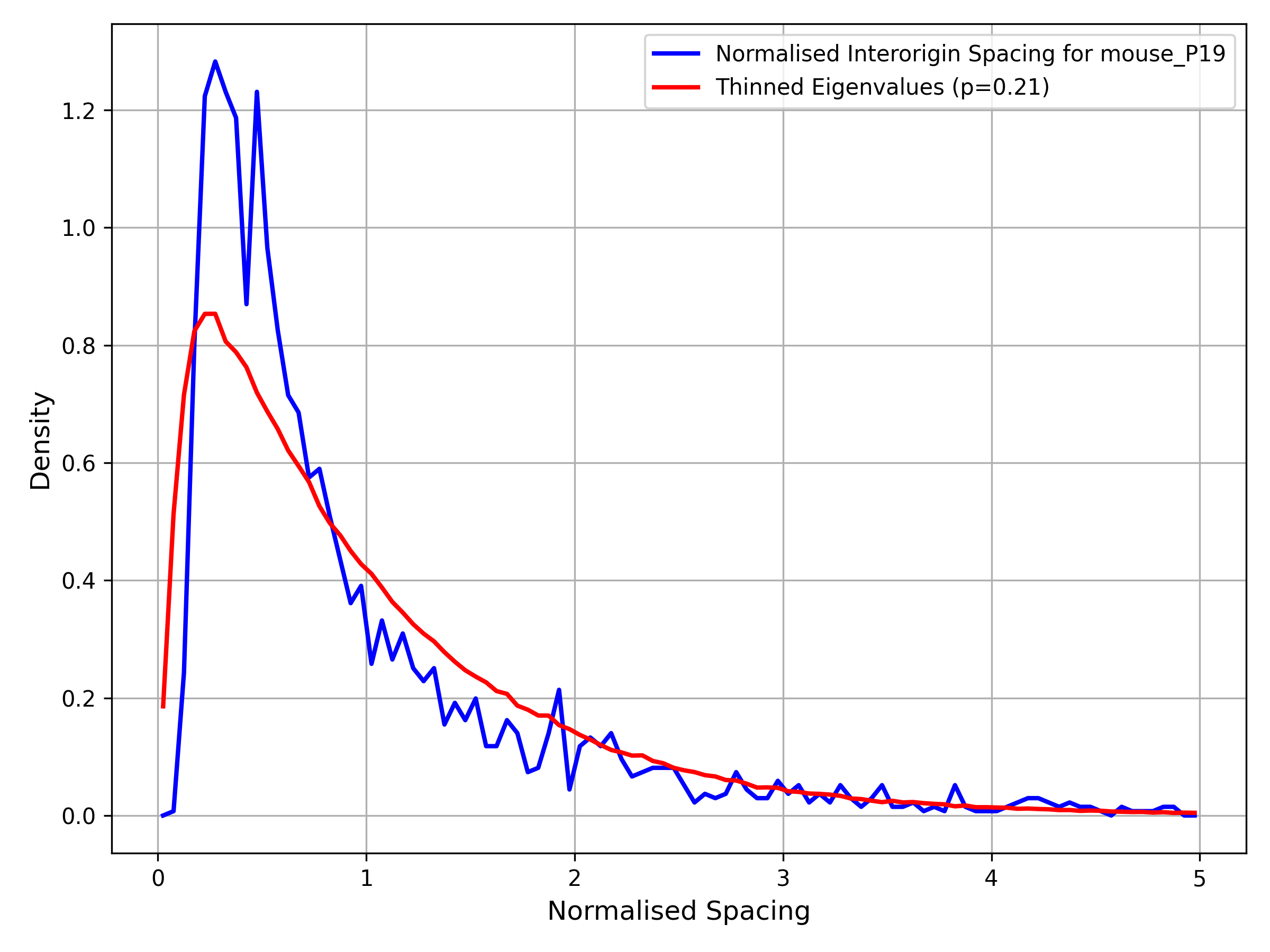} \includegraphics[scale=0.33]{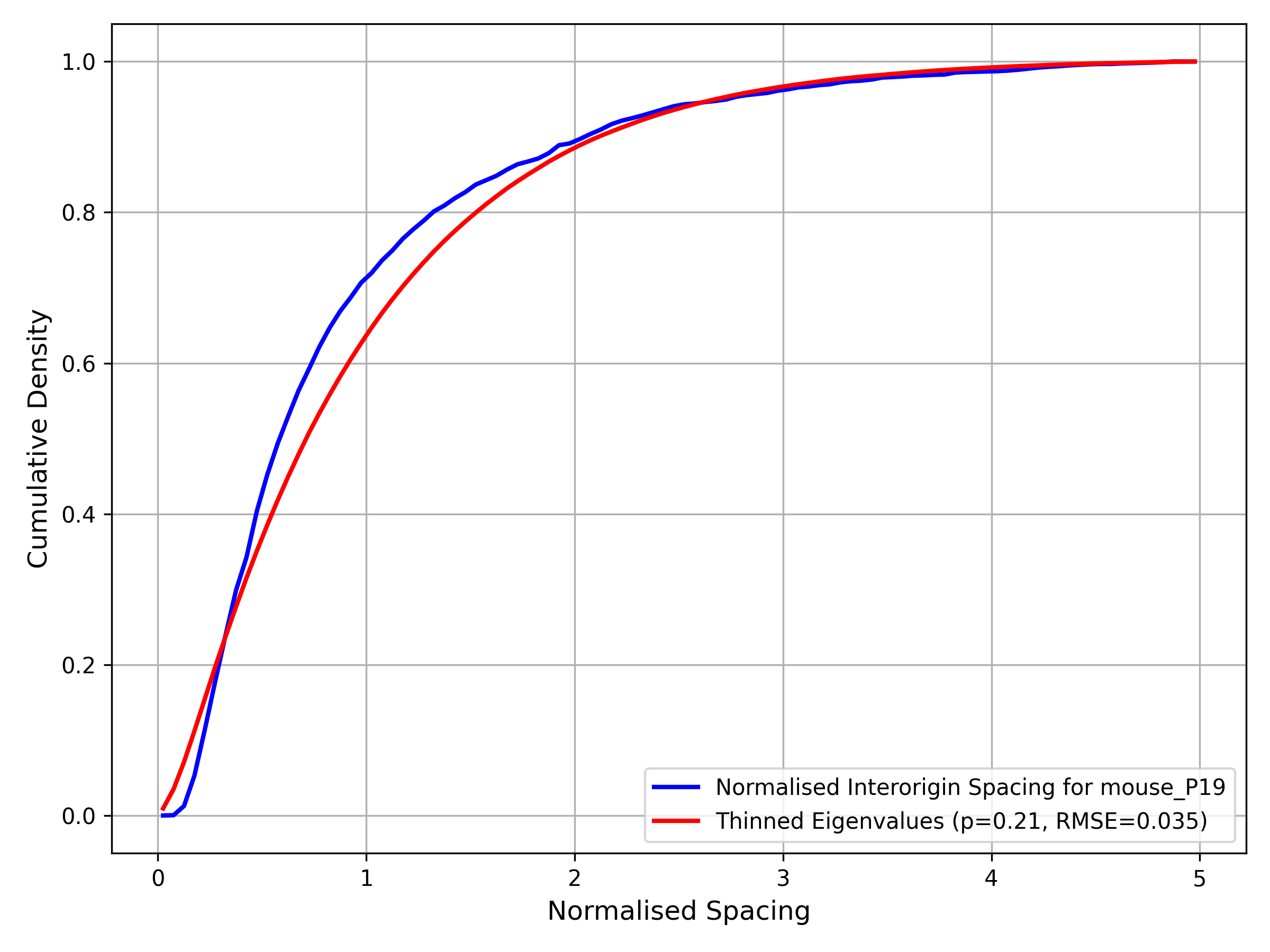}
    \caption{Left: Histogram of re-scaled spacings between midpoints of adjacent replication origins from mouse embryonic cells (or Mouse P19), data taken from \cite{Drosophila}, versus thinned COE eigenvalues with parameter $p=0.21$. Right: Corresponding cumulative distributions with RMSE of 0.035. Total Number of Spacings: 2747. Number of Chromosomes: 1.}
    \label{fig:MouseP19BestThin}
\end{figure}

Mouse embryonic cells (or Mouse ES1) as shown in Figure \ref{fig:MouseES1BestThin} are characterised by rapid growth rate, ease of DNA transfection (artificial insertion of DNA into cells), and clonability \cite{mouseES}.

Mouse Embryonic Fibroblasts (or Mouse MEF), in Figure \ref{fig:MouseMEFBestThin}, are a type of fibroblast prepared from mouse embryo. More detail on their characteristics and preparation can be found in \cite{xu2005preparation}.

Mouse teratocarcinoma cells (or Mouse P19) are the next dataset, as shown in Figure \ref{fig:MouseP19BestThin}. Teratocarcinoma is a form of malignant tumor that occurs in both animals and human \cite{lanza2005essentials} Essentially, we are looking at cells with some form of mutation, which still have some commonalities with genomes from other mouse cells \cite{Drosophila}.

In summary, we have an array of different types of cells from mice which still show similar origin spacing distribution, as seen in Figures \ref{fig:MouseES1BestThin}, \ref{fig:MouseMEFBestThin} and \ref{fig:MouseP19BestThin}. They are not a particularly good fit to any thinned COE ensemble.  The histograms show repulsion, but have a distribution that is more dominated by a peak of relatively small spacings than any distribution thinning COE eigenvalues can provide.

\begin{figure}[H]
    \centering
    \includegraphics[scale=0.33]{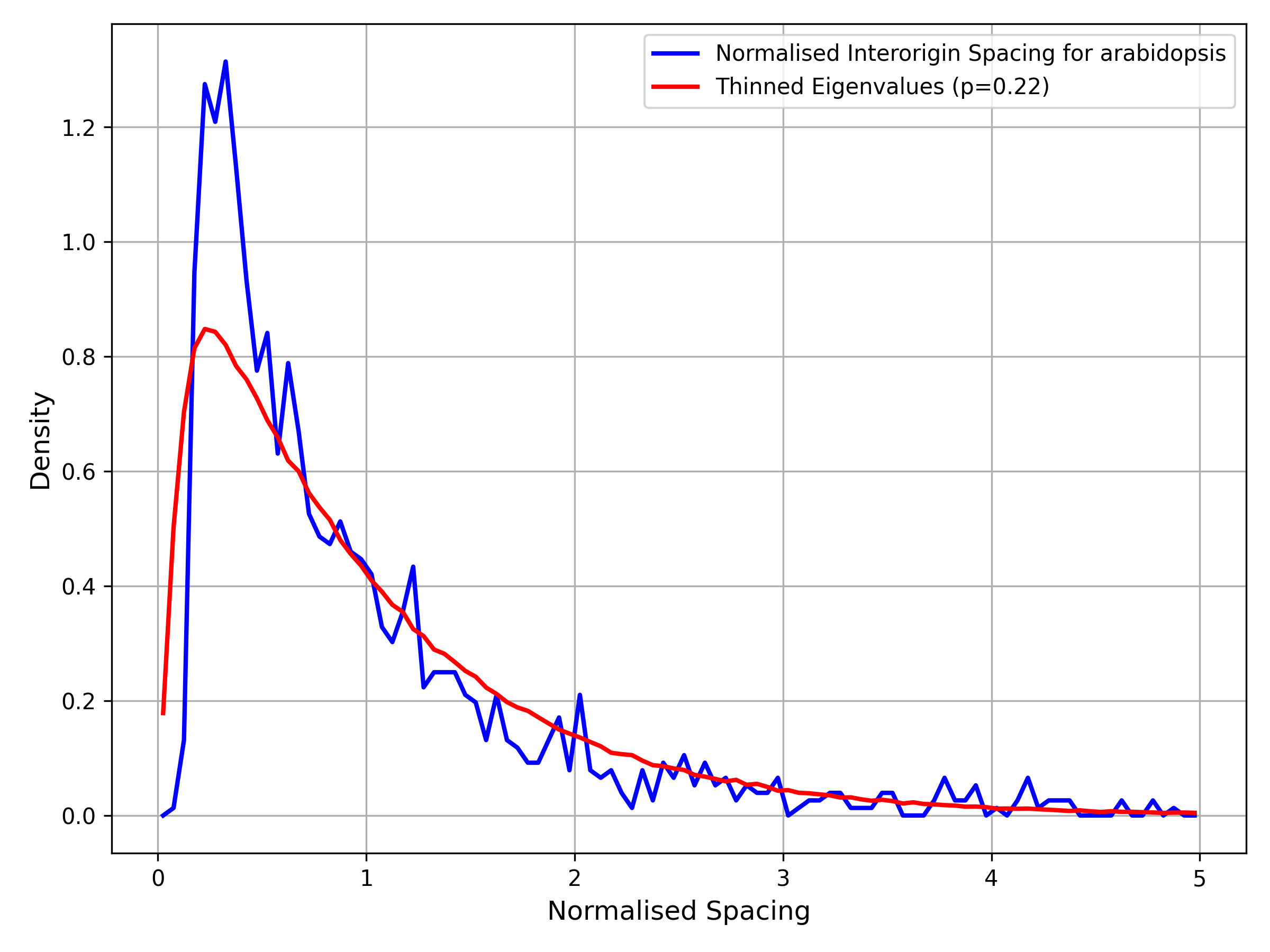} \includegraphics[scale=0.33]{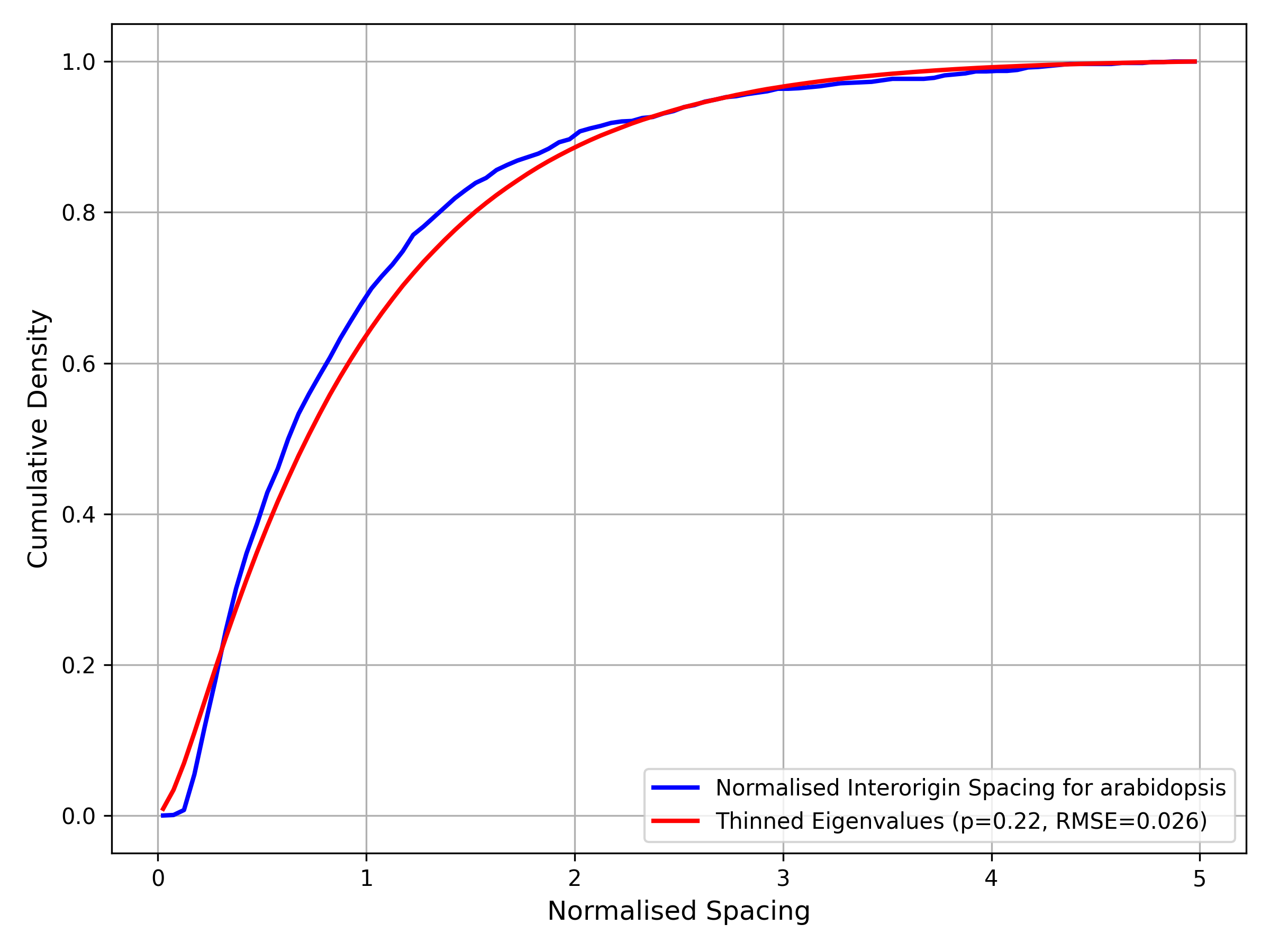}
    \caption{Left: Histogram of re-scaled spacings between midpoints of adjacent replication origins  from Arabidopsis thaliana (or just Arabidopsis), data taken from \cite{arabdopsis}, versus thinned COE eigenvalues with parameter $p=0.22$. Right: Corresponding cumulative distributions with RMSE of 0.026. Total Number of Spacings: 1538. Number of Chromosomes: 5.}
    \label{fig:ArabBestThin}
\end{figure}

Arabidopsis thaliana (or just Arabidopsis) is a small plant from the mustard family, it is actually considered a weed. It is considered a popular model organism in plant genetics. Despite the fact that it's a quite a complex multi-cellular organism, it has a relatively short genome. The original dataset and further analysis can be found in \cite{arabdopsis} and the plot is Figure \ref{fig:ArabBestThin}.

We see the distribution for Arabidopsis in Figure \ref{fig:ArabBestThin}. For plants such as Arabidopsis, DNA replication has similar constraints than in other eukaryotes but there are differences that lead to different to replication origin dynamics \cite{de2012regulating}. We see the same sort of behaviour as in the mouse data sets where there is a strong peak at relatively short spacings. 

\begin{figure}[H]
    \centering
    \includegraphics[scale=0.33]{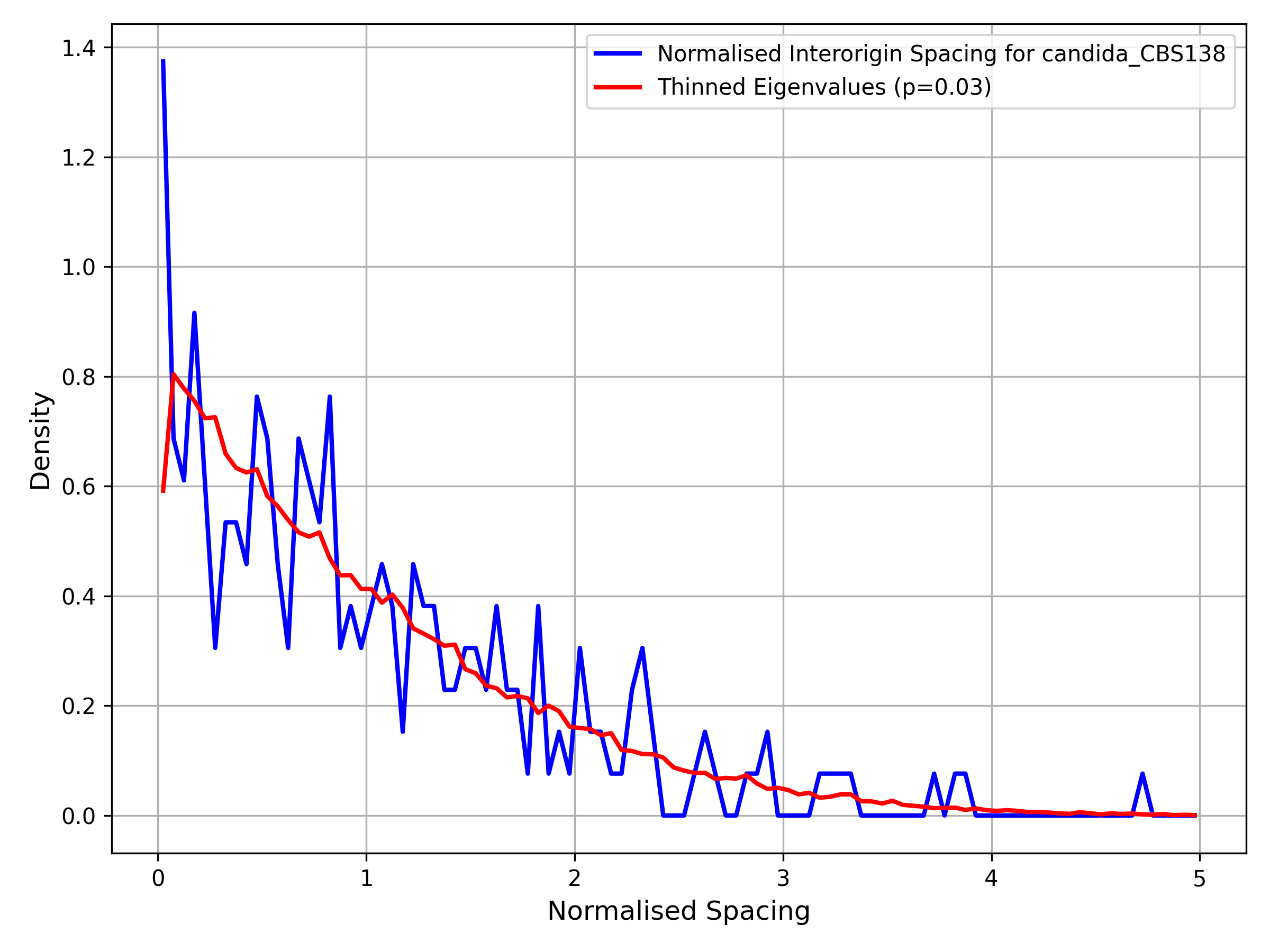} \includegraphics[scale=0.33]{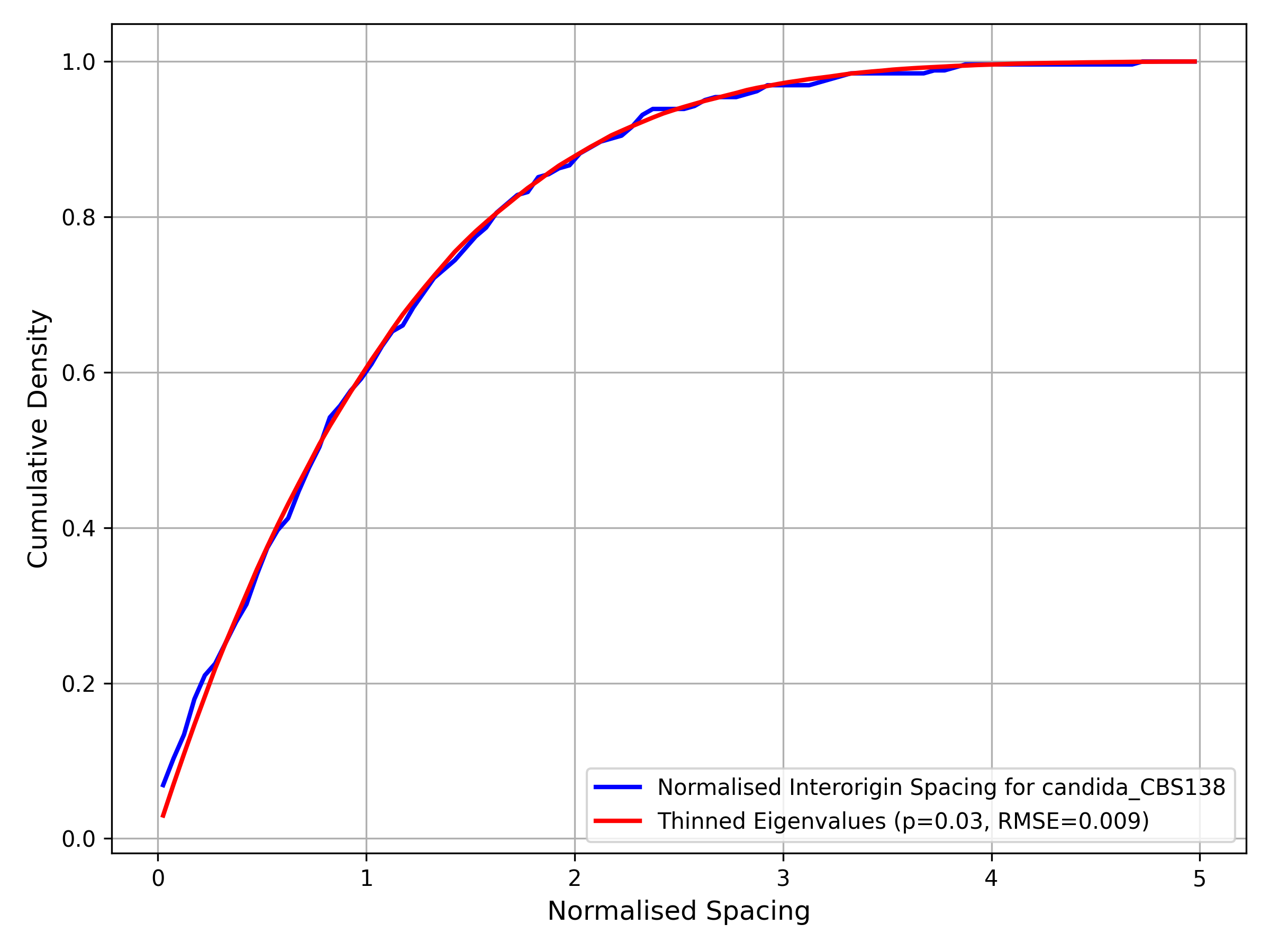}
    \caption{Left: Histogram of re-scaled spacings between midpoints of adjacent replication origins  from Candida glabrata (or Candida CBS138), data taken from \cite{candida1}, versus thinned COE eigenvalues with parameter $p=0.03$. Right: Corresponding cumulative distributions with RMSE of 0.009. Total Number of Spacings: 262. Number of  Chromosomes: 13.}
    \label{fig:CandidaBestThin}
\end{figure}

Candida glabrata (or Candida CBS138) is an asexual yeast strain closely related to S cerevisiae \cite{candida2}. It acts as an opportunistic pathogen which can cause candidiasis. The dataset has been taken from \cite{candida1} and can be seen in Figure \ref{fig:CandidaBestThin}.

Candida CBS138 is a relatively sparse dataset as shown in Figure \ref{fig:CandidaBestThin}, but shows almost perfectly uncorrelated spacings. A value of $p=0.03$ was the smallest we tested. Essentially this analysis tells us that the spacings between replication origins of Candida CBS138 are best modelled with an exponential random variable, showing completely uncorrelated spacings albeit for quite a small dataset.

We have looked at lots of different model organisms so far. The primary purpose of these model organisms is to serve as models for humans.  We have two human samples to look at.

Human K562 cells were from a 53-year-old female chronic myelogenous leukemia patient and taken from \cite{humank562} and Human MCF7 is a breast cancer cell line isolated in 1970 from a 69 year old White woman at the Michigan Cancer Foundation 7 (hence, MCF7). Original data from \cite{humanmcf7}.

Both of these datasets are characterised by a large number of origins across multiple chromosomes. 

\begin{figure}[H]
    \centering
    \includegraphics[scale=0.33]{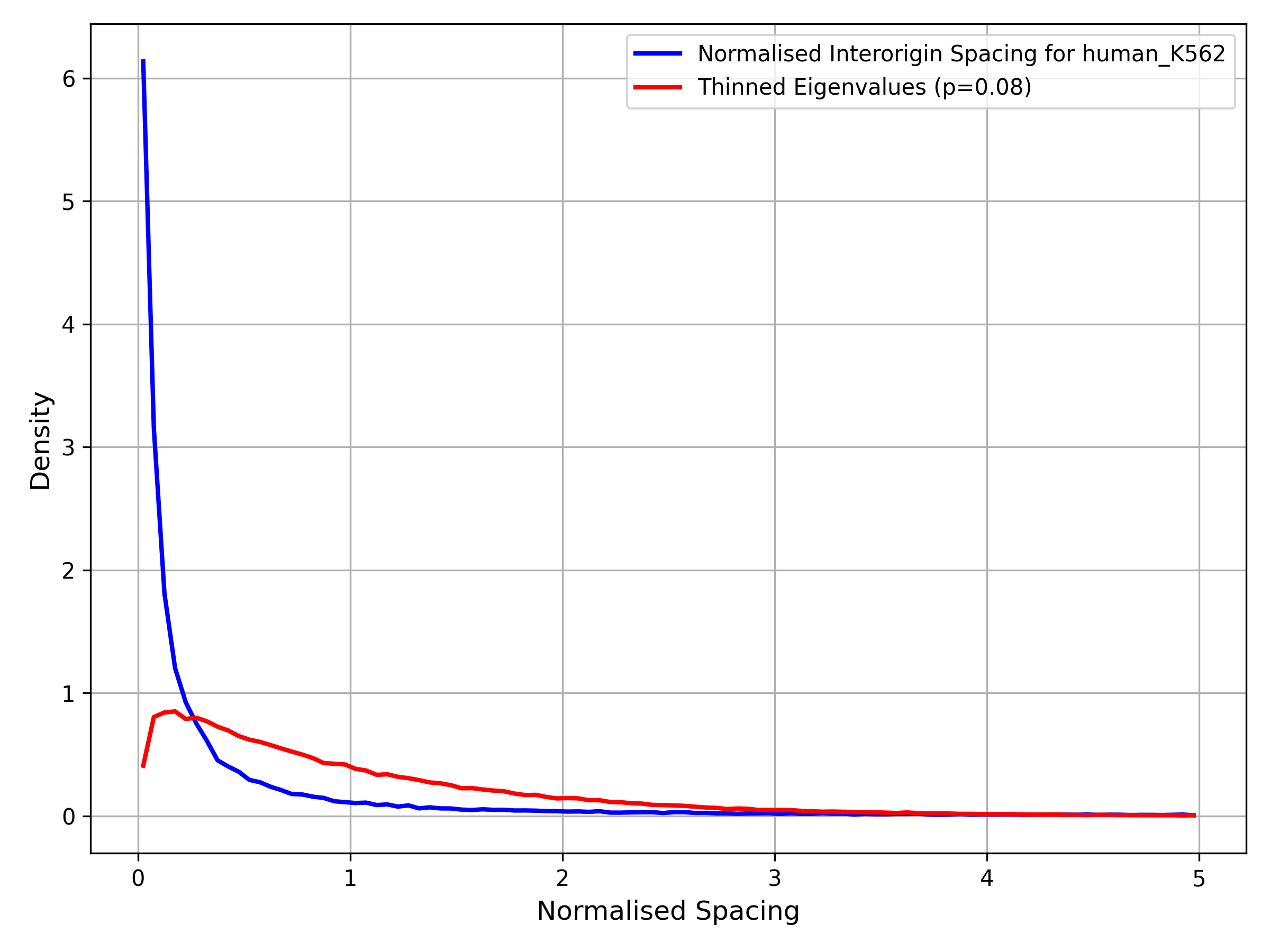} \includegraphics[scale=0.33]{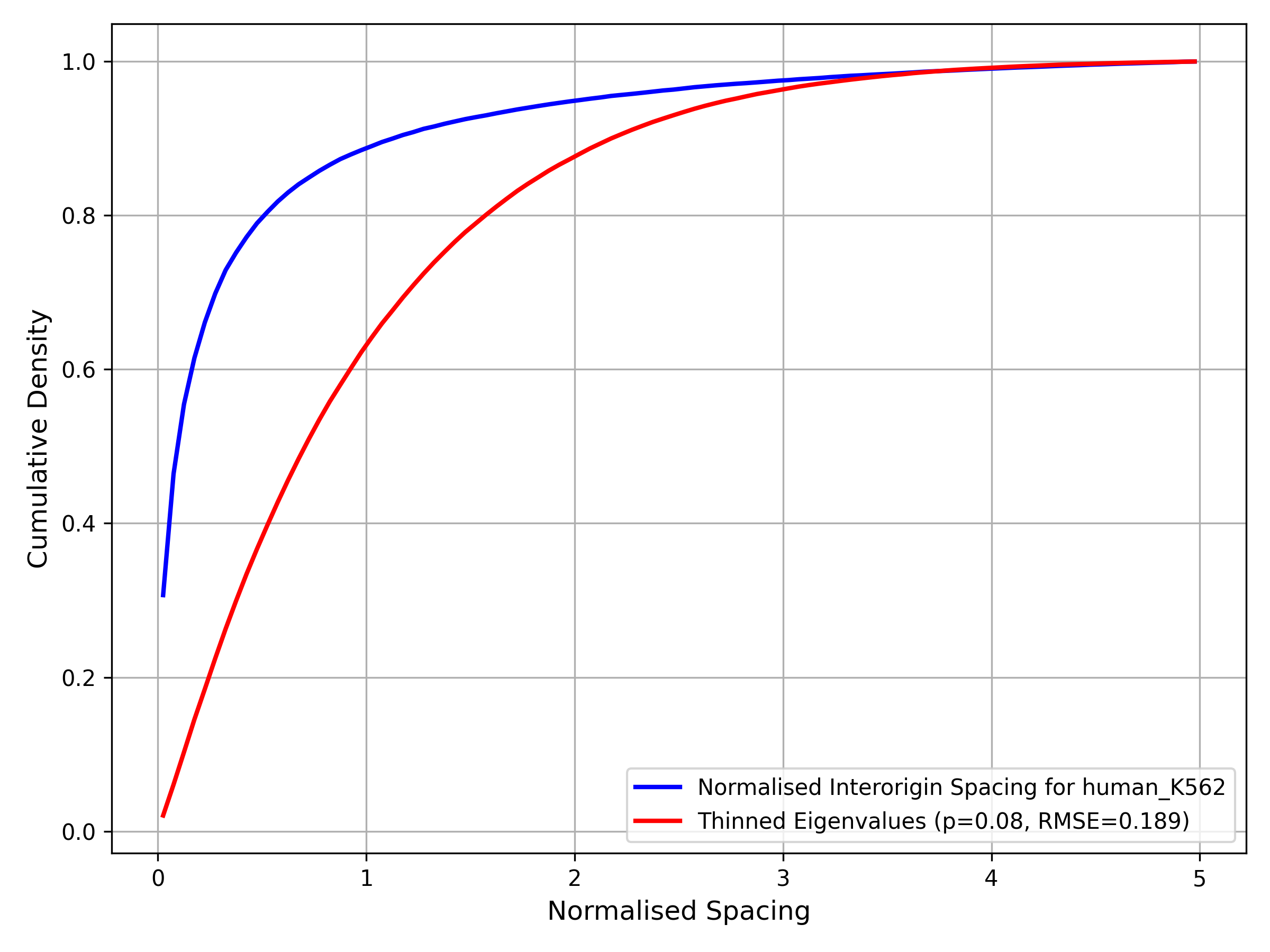}
    \caption{Left: Histogram of re-scaled spacings between midpoints of adjacent replication origins  from Human K562, data taken from \cite{humank562}, versus thinned COE eigenvalues with parameter $p=0.08$. Right: Corresponding cumulative distributions with RMSE of 0.189.  Total Number of Spacings: 62948. Number of Chromosomes: 23.}
    \label{fig:HumanK562BestThin}
\end{figure}

\begin{figure}[H]
    \centering
    \includegraphics[scale=0.33]{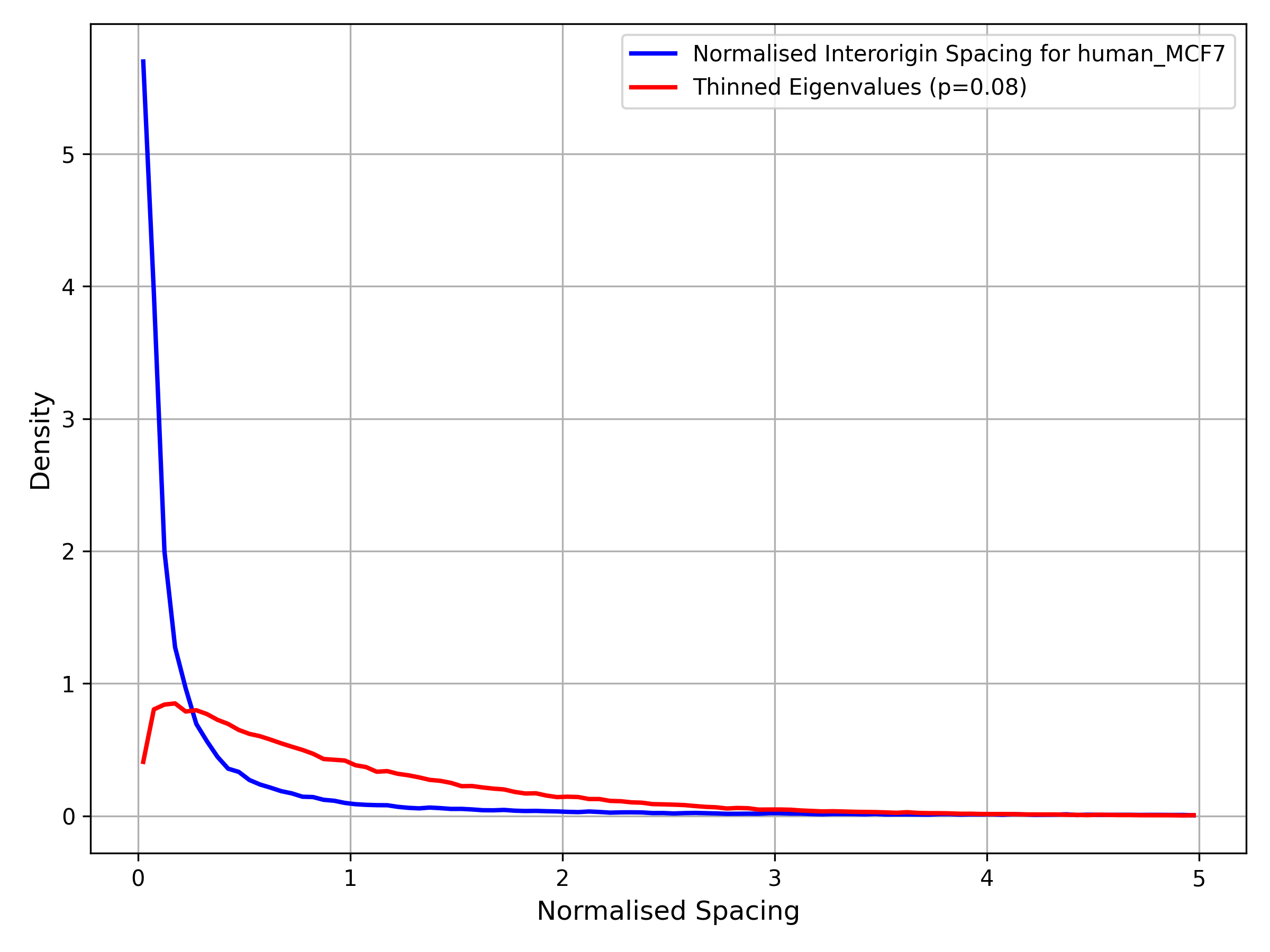} \includegraphics[scale=0.33]{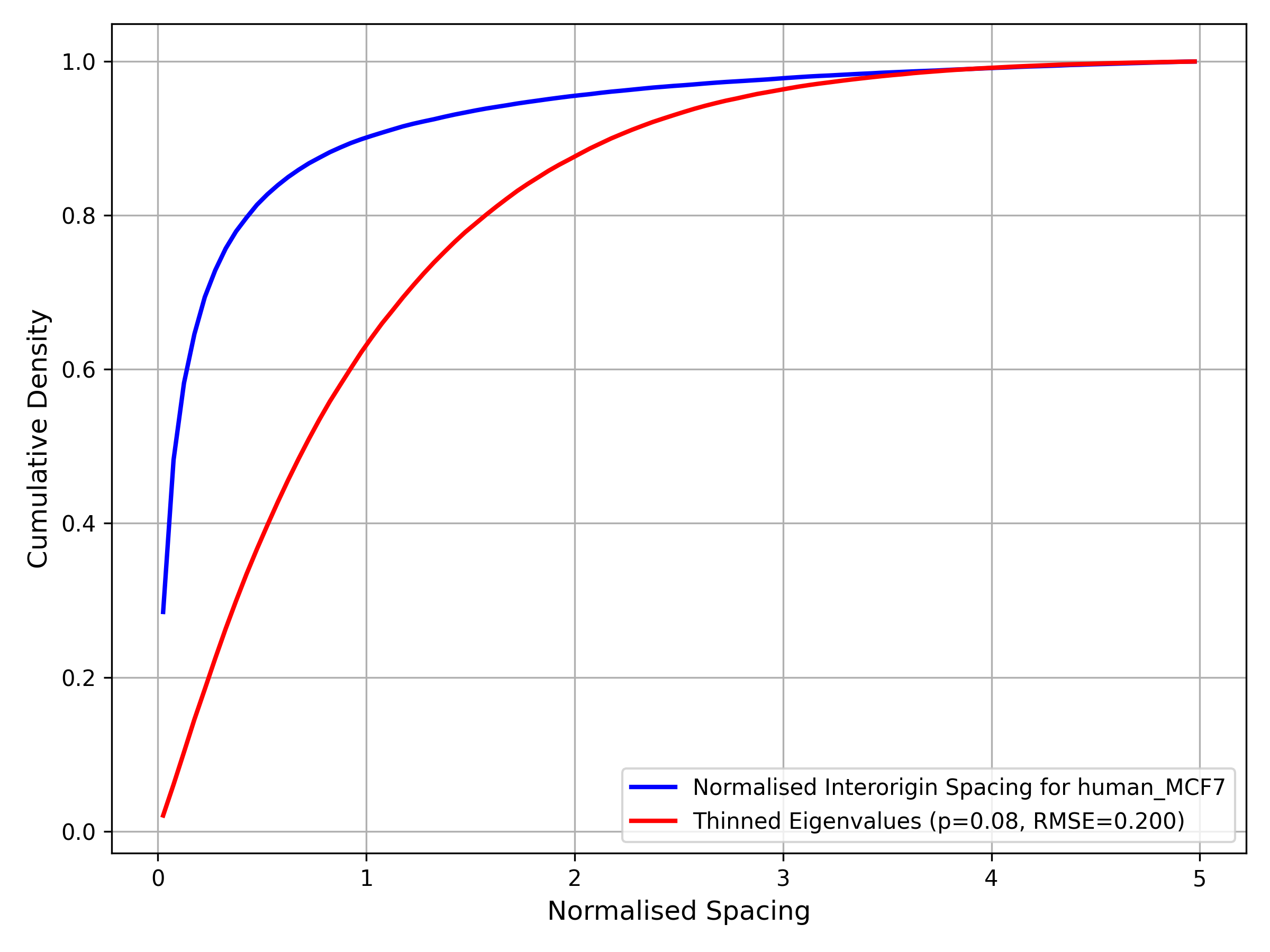}
    \caption{Left: Histogram of re-scaled spacings between midpoints of adjacent replication origins  from Human MCF7, data taken from \cite{humanmcf7}, versus thinned COE eigenvalues with parameter $p=0.08$. Right: Corresponding cumulative distributions with RMSE of 0.200.   Total Number of Spacings: 94172. Number of Chromosomes: 23.}
    \label{fig:HumanMCF7BestThin}
\end{figure}

In both Human datasets we see that a thinned COE eigenvalue spacing distribution is not a good fit for the data. The human datasets are too skewed to be effectively modelled by Wigner's surmise, an exponential random variable, or anything in between. These data sets are characterised by a number of very, very large spacings.  As we normalise the mean spacing to 1 for the plots, this has the effect of making the median spacing very small (less than 0.1 on the Human MCF7 plot in Figure \ref{fig:HumanMCF7BestThin}), even though more than 95\% of the spacings are less then twice the average spacing.  This implies there must be a very small number of very large spacings that are driving up the average spacing and causing the very steep slope near the origin on the cumulative spacing distribution after scaling by this large average value. 

Generally we observe that more complex organisms seem to have poorer fit to the random matrix models tested here, and seem to have a significant number of very large spacings and this would suggest that other mechanisms must have developed in those organisms so replication can succeed despite very large gaps between origins.

\begin{figure}[H]
	\centering
	\includegraphics[scale=0.6]{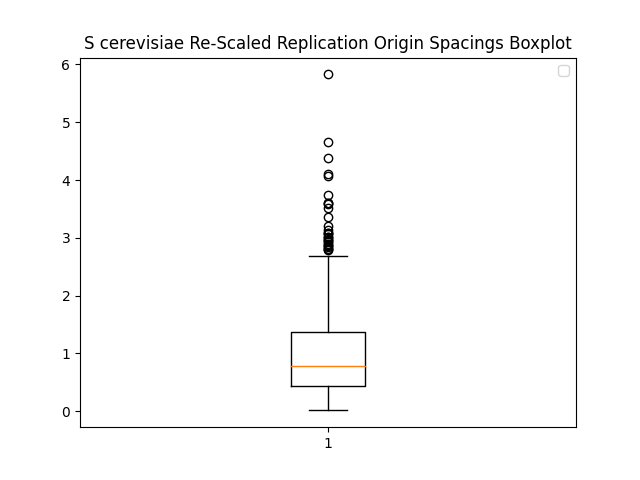}
	\caption{Box plot of the re-scaled spacings of the S. cerevisiae dataset as seen in Figure \ref{fig:ScereBestThin}. The $y$-axis represents the spacings with re-scaling  (i.e. the mean spacing is $1$). Outliers ($1.5$ times the interquartile range above or below the upper and lower quartiles) are marked as circles. Note that whilst there are outliers present, they are far less frequent and smaller than the outliers in Figures \ref{fig:HumanK562box} and \ref{fig:HumanMCF7box}.}
	\label{fig:ScereBox}
\end{figure}

\begin{figure}[H]
	\centering
	\includegraphics[scale=0.6]{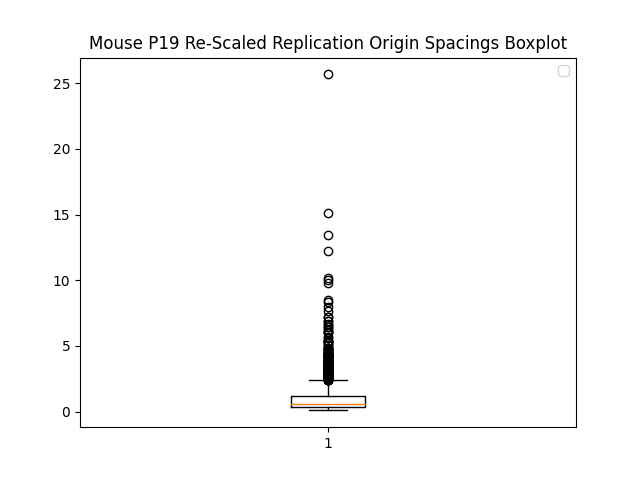}
	\caption{Box plot of the re-scaled spacings of the  Mouse P19 dataset as seen in Figure \ref{fig:MouseP19BestThin}.   The $y$-axis represents the spacings with re-scaling  (i.e. the mean spacing is $1$). Outliers ($1.5$ times the interquartile range above or below the upper and lower quartiles) are marked as circles. The outlier behaviour here seems to be between Figures \ref{fig:ScereBox} and \ref{fig:HumanK562box}.}
	\label{fig:mouseP19box}
\end{figure}

\begin{figure}[H]
	\centering
	\includegraphics[scale=0.6]{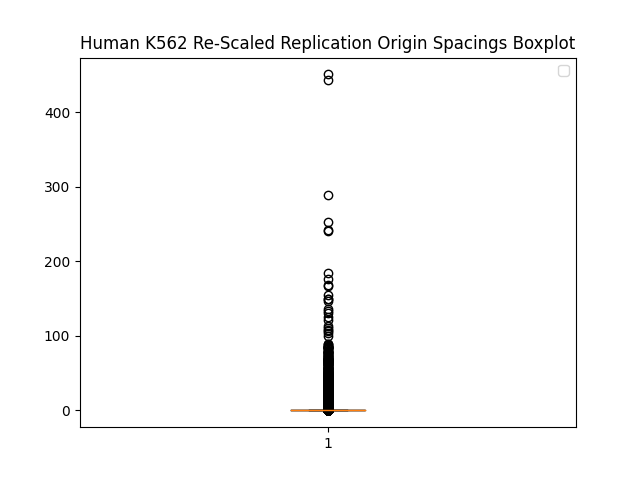}
	\caption{Box plot of the re-scaled spacings of the Human K562 dataset as seen in Figure \ref{fig:HumanK562BestThin}. The $y$-axis represents the spacings with re-scaling (i.e. the mean spacing is $1$). Outliers ($1.5$ times the interquartile range above or below the upper and lower quartiles) are marked as circles. The entire interquartile range here is visualised as a single line, indicating how extreme and prominent the outliers are in this dataset.}
	\label{fig:HumanK562box}
\end{figure}

\begin{figure}[H]
	\centering
	\includegraphics[scale=0.6]{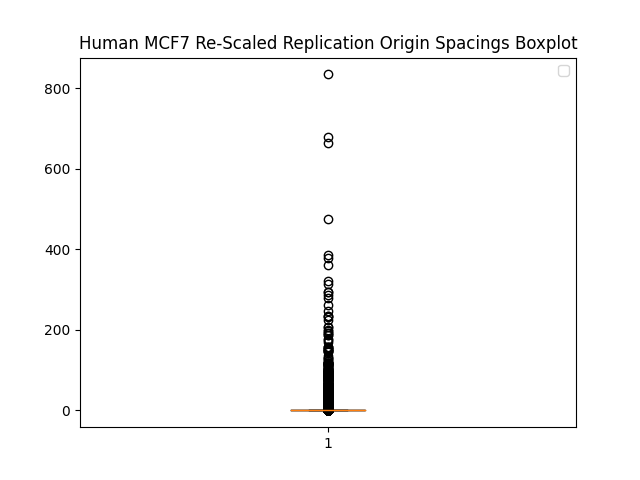}
	\caption{Box plot of the re-scaled spacings of the Human MCF7 dataset as seen in Figure \ref{fig:HumanMCF7BestThin}. The $y$-axis represents the spacings with re-scaling  (i.e. the mean spacing is $1$). Outliers ($1.5$ times the interquartile range above or below the upper and lower quartiles) are marked as circles. The entire interquartile range here is visualised as a single line, indicating how extreme and prominent the outliers are in this dataset.}
	\label{fig:HumanMCF7box}
\end{figure}

To visualise the quantity of extremely large spacings, we can look at the  outliers in box plots. We note that in Figures \ref{fig:ScereBox}, \ref{fig:mouseP19box}, \ref{fig:HumanK562box} and \ref{fig:HumanMCF7box} the mean spacing has been scaled to 1.

In Figure \ref{fig:ScereBox} of the S. cerevisiae data, we see that whilst there are outliers beyond 1.5 times the interquartile distance from the interquartile box, they are only a few multiples of the mean spacing, which is not out of line with a model like the thinned COE ensemble. 

In contrast, in Figures \ref{fig:HumanK562box} and \ref{fig:HumanMCF7box}, we see that the outliers are extremely significant, dwarfing interquartile range of the dataset in magnitude and appearing in relatively large frequencies compared to the rest of the data. It is evident that the two human datasets are heavily skewed by a significant amount of large outlier spacings. The same phenomenon, but to a far lesser extent, can also be seen in the mouse data, such as \ref{fig:mouseP19box}.

The range of different origin position statistics found across different organisms presumably reflects differing mechanisms at work in the process of DNA replication, with a general trend that the less complex organisms seem to have fewer very small or very large spaces between origins of replication. 

The boxplots for all of the other datasets can be found in Appendix A of \cite{kn:Day}.

\section*{Code}
All the code to produce simulations and plots for this work can be found on GitHub: \url{https://github.com/HuwWDay/RMTDNAData}

\section*{Acknowledgements}
We thank Beth Kent, Jack Simons, Lohini Sri Ram, Nor Farah Wahidah, Patrick Nairne, Sam Stockman, Zen Kok for their undergraduate projects on this topic. This gave us a head start which made approaching this problem considerably easier.

Thank you Ben Carter and Ellen Spackman for helping us understand the genetics content of this problem.

Thank you to M{\'a}rton Bal{\'a}zs for your help throughout this project.

\section*{Funding}

\textrm{The first author was supported and funded by the EPSRC, which is part of UKRI.}


\newcommand{\etalchar}[1]{$^{#1}$}

\end{document}